\RequirePackage{lineno}

\documentclass[aps,prl,twocolumn,showpacs,superscriptaddress,groupedaddress]{revtex4}

\usepackage{graphicx}
\usepackage{dcolumn} 
\usepackage{bm}      
\usepackage{amssymb} 
\usepackage{float} 
\usepackage{subfigure}

\newcommand{\comment}[1]{} 

\newcommand{\ppb}{\mbox{\ensuremath{p\bar p}}} 
 
\newcommand{\ttb}{\mbox{\ensuremath{t\bar t}}} 
\newcommand{\bbb}{\mbox{\ensuremath{b\bar b}}} 
\newcommand{\nunub}{\mbox{\ensuremath{\nu\bar\nu}}} 
 
\newcommand{\invfb}{fb$^{-1}$}

\newcommand{\met}{\mbox{\ensuremath{\not\!\!E_T}}}
\newcommand{\mht}{\mbox{\ensuremath{\not\!\!H_T}}}
\newcommand{\mpt}{\mbox{\ensuremath{\slash\kern-.5emp_{T}}}}
\newcommand{\et}{\mbox{\ensuremath{E_{T}}}}

\newcommand{\hht}{\mbox{\ensuremath{H_{T}}}}
\newcommand{\pt}{\mbox{\ensuremath{p_{T}}}}

\newcommand{\jj}{\mbox{\ensuremath{(\mathrm{jet}_1,\mathrm{jet}_2)}}}

\begin{document} 

\hspace{5.2in} \mbox{Fermilab-Pub-12-404-E}

\title{Search for the standard model Higgs boson in the $\boldsymbol{ZH\to\nunub\bbb}$ channel \\
 in 9.5~\invfb\ of $\boldsymbol{\ppb}$ collisions at $\boldsymbol{\sqrt{s}=1.96}$~TeV \\ }

\affiliation{LAFEX, Centro Brasileiro de Pesquisas F\'{i}sicas, Rio de Janeiro, Brazil}
\affiliation{Universidade do Estado do Rio de Janeiro, Rio de Janeiro, Brazil}
\affiliation{Universidade Federal do ABC, Santo Andr\'e, Brazil}
\affiliation{University of Science and Technology of China, Hefei, People's Republic of China}
\affiliation{Universidad de los Andes, Bogot\'a, Colombia}
\affiliation{Charles University, Faculty of Mathematics and Physics, Center for Particle Physics, Prague, Czech Republic}
\affiliation{Czech Technical University in Prague, Prague, Czech Republic}
\affiliation{Center for Particle Physics, Institute of Physics, Academy of Sciences of the Czech Republic, Prague, Czech Republic}
\affiliation{Universidad San Francisco de Quito, Quito, Ecuador}
\affiliation{LPC, Universit\'e Blaise Pascal, CNRS/IN2P3, Clermont, France}
\affiliation{LPSC, Universit\'e Joseph Fourier Grenoble 1, CNRS/IN2P3, Institut National Polytechnique de Grenoble, Grenoble, France}
\affiliation{CPPM, Aix-Marseille Universit\'e, CNRS/IN2P3, Marseille, France}
\affiliation{LAL, Universit\'e Paris-Sud, CNRS/IN2P3, Orsay, France}
\affiliation{LPNHE, Universit\'es Paris VI and VII, CNRS/IN2P3, Paris, France}
\affiliation{CEA, Irfu, SPP, Saclay, France}
\affiliation{IPHC, Universit\'e de Strasbourg, CNRS/IN2P3, Strasbourg, France}
\affiliation{IPNL, Universit\'e Lyon 1, CNRS/IN2P3, Villeurbanne, France and Universit\'e de Lyon, Lyon, France}
\affiliation{III. Physikalisches Institut A, RWTH Aachen University, Aachen, Germany}
\affiliation{Physikalisches Institut, Universit\"at Freiburg, Freiburg, Germany}
\affiliation{II. Physikalisches Institut, Georg-August-Universit\"at G\"ottingen, G\"ottingen, Germany}
\affiliation{Institut f\"ur Physik, Universit\"at Mainz, Mainz, Germany}
\affiliation{Ludwig-Maximilians-Universit\"at M\"unchen, M\"unchen, Germany}
\affiliation{Fachbereich Physik, Bergische Universit\"at Wuppertal, Wuppertal, Germany}
\affiliation{Panjab University, Chandigarh, India}
\affiliation{Delhi University, Delhi, India}
\affiliation{Tata Institute of Fundamental Research, Mumbai, India}
\affiliation{University College Dublin, Dublin, Ireland}
\affiliation{Korea Detector Laboratory, Korea University, Seoul, Korea}
\affiliation{CINVESTAV, Mexico City, Mexico}
\affiliation{Nikhef, Science Park, Amsterdam, the Netherlands}
\affiliation{Radboud University Nijmegen, Nijmegen, the Netherlands}
\affiliation{Joint Institute for Nuclear Research, Dubna, Russia}
\affiliation{Institute for Theoretical and Experimental Physics, Moscow, Russia}
\affiliation{Moscow State University, Moscow, Russia}
\affiliation{Institute for High Energy Physics, Protvino, Russia}
\affiliation{Petersburg Nuclear Physics Institute, St. Petersburg, Russia}
\affiliation{Instituci\'{o} Catalana de Recerca i Estudis Avan\c{c}ats (ICREA) and Institut de F\'{i}sica d'Altes Energies (IFAE), Barcelona, Spain}
\affiliation{Uppsala University, Uppsala, Sweden}
\affiliation{Lancaster University, Lancaster LA1 4YB, United Kingdom}
\affiliation{Imperial College London, London SW7 2AZ, United Kingdom}
\affiliation{The University of Manchester, Manchester M13 9PL, United Kingdom}
\affiliation{University of Arizona, Tucson, Arizona 85721, USA}
\affiliation{University of California Riverside, Riverside, California 92521, USA}
\affiliation{Florida State University, Tallahassee, Florida 32306, USA}
\affiliation{Fermi National Accelerator Laboratory, Batavia, Illinois 60510, USA}
\affiliation{University of Illinois at Chicago, Chicago, Illinois 60607, USA}
\affiliation{Northern Illinois University, DeKalb, Illinois 60115, USA}
\affiliation{Northwestern University, Evanston, Illinois 60208, USA}
\affiliation{Indiana University, Bloomington, Indiana 47405, USA}
\affiliation{Purdue University Calumet, Hammond, Indiana 46323, USA}
\affiliation{University of Notre Dame, Notre Dame, Indiana 46556, USA}
\affiliation{Iowa State University, Ames, Iowa 50011, USA}
\affiliation{University of Kansas, Lawrence, Kansas 66045, USA}
\affiliation{Kansas State University, Manhattan, Kansas 66506, USA}
\affiliation{Louisiana Tech University, Ruston, Louisiana 71272, USA}
\affiliation{Boston University, Boston, Massachusetts 02215, USA}
\affiliation{Northeastern University, Boston, Massachusetts 02115, USA}
\affiliation{University of Michigan, Ann Arbor, Michigan 48109, USA}
\affiliation{Michigan State University, East Lansing, Michigan 48824, USA}
\affiliation{University of Mississippi, University, Mississippi 38677, USA}
\affiliation{University of Nebraska, Lincoln, Nebraska 68588, USA}
\affiliation{Rutgers University, Piscataway, New Jersey 08855, USA}
\affiliation{Princeton University, Princeton, New Jersey 08544, USA}
\affiliation{State University of New York, Buffalo, New York 14260, USA}
\affiliation{University of Rochester, Rochester, New York 14627, USA}
\affiliation{State University of New York, Stony Brook, New York 11794, USA}
\affiliation{Brookhaven National Laboratory, Upton, New York 11973, USA}
\affiliation{Langston University, Langston, Oklahoma 73050, USA}
\affiliation{University of Oklahoma, Norman, Oklahoma 73019, USA}
\affiliation{Oklahoma State University, Stillwater, Oklahoma 74078, USA}
\affiliation{Brown University, Providence, Rhode Island 02912, USA}
\affiliation{University of Texas, Arlington, Texas 76019, USA}
\affiliation{Southern Methodist University, Dallas, Texas 75275, USA}
\affiliation{Rice University, Houston, Texas 77005, USA}
\affiliation{University of Virginia, Charlottesville, Virginia 22904, USA}
\affiliation{University of Washington, Seattle, Washington 98195, USA}
\author{V.M.~Abazov} \affiliation{Joint Institute for Nuclear Research, Dubna, Russia}
\author{B.~Abbott} \affiliation{University of Oklahoma, Norman, Oklahoma 73019, USA}
\author{B.S.~Acharya} \affiliation{Tata Institute of Fundamental Research, Mumbai, India}
\author{M.~Adams} \affiliation{University of Illinois at Chicago, Chicago, Illinois 60607, USA}
\author{T.~Adams} \affiliation{Florida State University, Tallahassee, Florida 32306, USA}
\author{G.D.~Alexeev} \affiliation{Joint Institute for Nuclear Research, Dubna, Russia}
\author{G.~Alkhazov} \affiliation{Petersburg Nuclear Physics Institute, St. Petersburg, Russia}
\author{A.~Alton$^{a}$} \affiliation{University of Michigan, Ann Arbor, Michigan 48109, USA}
\author{G.~Alverson} \affiliation{Northeastern University, Boston, Massachusetts 02115, USA}
\author{A.~Askew} \affiliation{Florida State University, Tallahassee, Florida 32306, USA}
% ...
%
%
\collaboration{The D0 Collaboration\footnote{with visitors from
$^{a}$Augustana College, Sioux Falls, SD, USA,
$^{b}$The University of Liverpool, Liverpool, UK,
$^{c}$UPIITA-IPN, Mexico City, Mexico,
$^{d}$DESY, Hamburg, Germany,
$^{e}$SLAC, Menlo Park, CA, USA,
$^{f}$University College London, London, UK,
$^{g}$Centro de Investigacion en Computacion - IPN, Mexico City, Mexico,
$^{h}$ECFM, Universidad Autonoma de Sinaloa, Culiac\'an, Mexico
and
$^{i}$Universidade Estadual Paulista, S\~ao Paulo, Brazil.
}} \noaffiliation
\vskip 0.25cm

\date{July 24, 2012}

\begin{abstract}
We present a search for the standard model Higgs boson in 9.5~\invfb\ 
of \ppb\ collisions at $\sqrt{s}=1.96$~TeV 
collected with the D0 detector at the Fermilab Tevatron Collider. 
The final state considered contains a pair of $b$ jets and is characterized by an imbalance in
transverse energy, 
as expected from $\ppb\to ZH\to\nunub\bbb$ production. The search is also 
sensitive to the $WH\to\ell\nu\bbb$ channel when the charged lepton is not 
identified. The data are found to be in good agreement with the expected background.
For a Higgs boson mass of 125~GeV, we set a limit at the 
95\% C.L. on the cross section $\sigma(\ppb\to [Z/W]H)$, assuming standard model
branching fractions, that is a factor of 4.3 times larger than the
theoretical standard model value, while the expected factor is 3.9.
The search is also used to measure a combined $WZ$ and $ZZ$ production cross section that is a factor of 
$0.94 \pm 0.31\thinspace\mathrm{(stat)} \pm 0.34\thinspace\mathrm{(syst)}$ 
times the standard model prediction of 4.4~pb, 
with an observed significance of 2.0 standard deviations. 
\end{abstract}

\maketitle 

\section{Introduction}\label{sec:intro}

In the standard model (SM)~\cite{GWS}, electroweak symmetry breaking is achieved via the
introduction of a doublet of scalar fields, of which one degree of freedom
remains once the $W$ and $Z$ vector bosons have acquired their masses. This degree
of freedom manifests itself as a new scalar particle~\cite{HBEGHK}, the Higgs boson ($H$).
Associated $ZH$ production in \ppb\ collisions at $\sqrt{s}=1.96$~TeV, 
with $Z\to\nunub$ and $H\to\bbb$, is among the most sensitive processes in the search for
a Higgs boson with mass $M_H \lesssim 135$~GeV at the Fermilab Tevatron 
Collider~\cite{prospects}.
The D0 Collaboration published a search for this process based on 
5.2~\invfb\ of integrated luminosity~\cite{vvbbPubl}. 
In this Letter, an extension of this search to the full Run~II dataset 
is presented. 
The CDF Collaboration recently reported results from a similar search~\cite{CDFvvbb},
as well as the ATLAS and CMS Collaborations using $pp$ collisions at 7 TeV at the 
LHC~\cite{CMS_bb,ATLAS_bb}.
A lower limit of 114.4~GeV was set on $M_H$ by the LEP Collaborations~\cite{LEPH}, 
while an upper limit at 127 GeV has been established by the ATLAS and CMS 
Collaborations~\cite{CMS_combo,ATLAS_combo}.
These limits and those given below are all defined at the 95\% C.L. 
The ATLAS and CMS Collaborations have also published~\cite{CMS_combo,ATLAS_combo} 
excesses above background expectations at approximately 125 GeV and have recently 
reported results confirming these excesses at the five standard deviations 
level~\cite{ATLAS_observ,CMS_observ}.

The final-state topology considered in this search
consists of a pair of $b$ jets from $H\to\bbb$ decay and missing
transverse energy (\met) from $Z\to\nunub$.  The search is 
also sensitive to the $WH$ process when the charged lepton from
$W\to\ell\nu$ decay is not identified. The main backgrounds arise from
$(W/Z)$+heavy-flavor jets (jets initiated by $b$ or $c$ quarks), top quark
production, and multijet (MJ) events with \met\ arising from
mismeasurement of jet energies.
A boosted-decision-tree discriminant based on kinematic properties 
is first used to reject most of the multijet events.
Next, jets from candidate Higgs boson decays are required to be 
identified as $b$ jets. Finally, 
discrimination between signal and remaining backgrounds is achieved by 
means of additional boosted decision trees.

To validate the techniques used in the search for the Higgs boson, 
the analysis is also interpreted as a measurement of $WZ$ and $ZZ$ diboson production. 
The only modification is in the training of the final discriminants, 
for which a diboson signal is used instead of a Higgs boson signal.

\section{Data and simulated samples}

The D0 detector used for Tevatron Run II (2001 -- 2011)
is described in detail in Ref.~\cite{Dzero}. Its main components are:
a tracking system surrounding the beam pipe, followed by
a liquid-argon and uranium sampling calorimeter, and then a muon system. 
The tracking system is immersed in a 2~T magnetic field provided by 
a superconducting solenoid and consists of a silicon microstrip tracker 
followed by a scintillating fiber tracker. The calorimeter is composed of 
a central and two end sections housed in separate cryostats. Each section is segmented
in depth, with four electromagnetic layers followed by up to five hadronic layers.
Scintillating tiles provide additional sampling between the cryostats. The
muon system consists of tracking and trigger detectors in front of and beyond
1.8~T iron toroids. Online event selection is provided by a three-level trigger system.

The data used in this
analysis were recorded using triggers designed to select events with
jets and \met~\cite{ochando}. After imposing data quality requirements, the total integrated
luminosity recorded with these triggers is 9.5~\invfb, corresponding to all available Run II 
data for this analysis. 

The analysis relies on (i) charged particle
tracks, (ii) calorimeter jets reconstructed in a cone of radius 0.5 in $y$-$\phi$
space, where $y$ is the rapidity and $\phi$ the azimuthal angle,
using the iterative midpoint cone algorithm~\cite{jetalgo}, and (iii)
electrons or muons identified through the association of tracks with
electromagnetic calorimeter clusters or with hits in the muon
detector, respectively.  The \met\ is reconstructed as the negative of
the vectorial sum of the transverse components of energy deposits in the
calorimeter and is corrected for identified muons. Jet energies are
calibrated using primarily transverse energy balance in photon+jet events~\cite{mikko}, 
and these corrections are propagated to the \met\ assessment.

Those backgrounds arising from MJ processes with instrumental 
effects giving rise to \met\ are estimated from data. 
The remainder of the backgrounds and the signal processes are simulated 
by Monte Carlo (MC).
Events from $(W/Z)$+jets processes are generated with 
{\sc alpgen}~\cite{alpgen}, interfaced with {\sc pythia}~\cite{pythia} for 
initial- and final-state radiation and for hadronization. 
The \pt\ spectrum of the $Z$ boson is reweighted to match the D0 
measurement~\cite{Zpt}. The \pt\ spectrum of the $W$ boson is reweighted 
using the same experimental input, corrected for the differences between 
the $W$ and $Z$ \pt\ spectra predicted in next-to-next-to-leading order 
(NNLO) QCD~\cite{MelPet}.
To simulate \ttb\ and electroweak single top quark production, the {\sc alpgen} and 
{\sc singletop}~\cite{comphep} generators, respectively, 
are interfaced with {\sc pythia}, while 
vector boson pair production is generated with {\sc pythia}. The 
$ZH$ and $WH$ signal processes are generated with {\sc pythia} 
for Higgs boson masses from 100 to 150 GeV in 5 GeV steps. 
All these simulations use CTEQ6L1 parton distribution functions 
(PDFs)~\cite{cteq}.

The absolute normalizations for $(W/Z)$ inclusive production are obtained from
NNLO calculations of total cross sections~\cite{WZcross} using the MSTW2008 NNLO PDFs~\cite{MRST}.
The heavy-flavor fractions in $(W/Z)$+jets are obtained using {\sc mcfm}~\cite{mcfm}
at next-to-leading order (NLO). The diboson cross sections are also
calculated with {\sc mcfm}~\cite{mcfmdiboson}. Cross sections for pair and single top
quark production are taken from Ref.~\cite{xsections}. 
For signal processes, cross sections are taken from Ref.~\cite{signal}.

Signal and background samples are passed through a full 
{\sc geant3}-based simulation~\cite{geant} of the detector 
response and processed with the same reconstruction program as used for data. 
Events from randomly selected beam crossings 
with the same instantaneous luminosity distribution as data are overlaid on simulated 
events to account for detector noise and 
contributions from additional \ppb\ interactions. 
Parameterizations of the trigger efficiencies are determined using events collected 
with independent triggers 
based on information from the muon detectors. Corrections for residual differences between data 
and simulation are applied for electron, muon, and jet 
identification. Jet energy calibration and resolution are adjusted in
simulated events to match those measured in data.

\section{Event selection}

A preselection that greatly reduces the overwhelming background from
multijet events is performed as follows.  The interaction vertex must be
reconstructed within the acceptance of the silicon vertex detector and at least three 
tracks must originate from that vertex. Jets with
associated tracks that meet criteria ensuring that
the $b$-tagging algorithm operates efficiently are denoted as ``taggable'' jets, 
except for those also identified as hadronic decays of $\tau$ leptons~\cite{tauID}.
Exactly two taggable jets are required, one of which must be the leading 
(highest \pt) jet
in the event; the Higgs candidate is formed from these two jets, 
denoted $\mathrm{jet_1}$ and $\mathrm{jet_2}$ (ordered in decreasing \pt). 
These jets must have transverse momentum $\pt>20$~GeV and
pseudorapidity $\vert\eta\vert<2.5$.
The two taggable jets 
must not be back-to-back in the plane transverse to the
beam direction: $\Delta\phi\jj<165^\circ$. Finally,~~$\met>40$~GeV is required.

Additional selection criteria define four distinct samples: 
(i) an ``analysis'' sample used to search for a Higgs boson signal; 
(ii) an ``electroweak (EW) control'' sample used to validate the background MC simulation,
enriched in $W(\to\mu\nu)$+jets events where the 
jet system has a topology similar to that of the analysis sample; 
(iii) an ``MJ-model'' sample, dominated by multijet events, used to 
model the MJ background in the analysis sample; and
(iv) a large ``MJ-enriched'' sample, used to validate this MJ-modeling 
procedure.  

The analysis sample is selected by requiring
the scalar sum of the transverse momenta of the two leading taggable jets to be 
greater than 80~GeV
and a measure of the \met\ significance ${\cal S} > 5$~\cite{metsig}. 
Larger values of $\cal S$ correspond to \met\ values 
that are less likely to be caused by fluctuations in jet 
energies. The $\cal S$ distribution is shown for the analysis sample
in Fig.\ \ref{HM_analysis_pretag}.

\begin{figure}[htp]
\includegraphics[width=8.5cm]{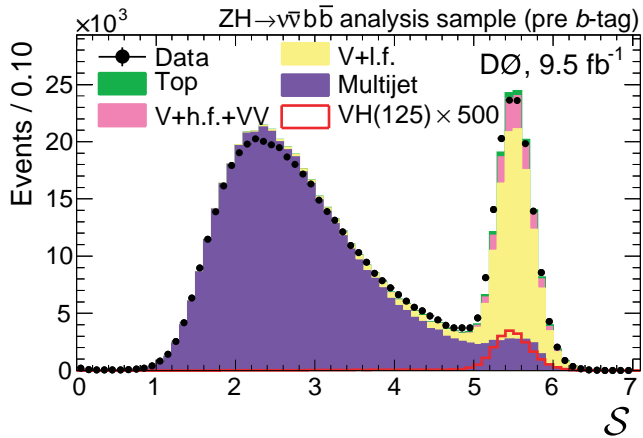}
\caption{\label{HM_analysis_pretag}  
(Color online.) The measure ${\cal S}$ of \met\ significance in the analysis sample 
without the requirement that ${\cal S}$ be larger than 5.
The data are shown as points with error bars and the background contributions as histograms: 
dibosons are labeled as ``VV,'' ``V+l.f.''\ includes $(W/Z)$+$(u,d,s,g)$ jets, 
``V+h.f.''\ includes $(W/Z)$+$(b,c)$ jets, 
and ``Top'' includes pair and single top quark production.
The distribution for signal (VH) is scaled by a factor of 500 and includes 
$ZH$ and $WH$ production for $M_H=125$~GeV.
}
\end{figure}

The dominant signal topology is a pair of $b$ jets recoiling
against the \met\ due to the neutrinos from $Z\to\nu\bar\nu$ decay,
leading to the direction of the \met\ being at a large angle with respect to the 
direction of each jet. In contrast, in events from MJ background with 
fluctuations in jet energy measurement, the \met\ tends to be aligned
with a mismeasured jet. A second estimate of the \met\ can be obtained from the 
missing $p_T$, $\mpt$, calculated from the reconstructed charged particle tracks 
originating from the interaction vertex.
This variable is less sensitive to jet energy measurement fluctuations.
In signal events, \mpt\ is also expected to point away from both
jets, while for MJ background, its angular distribution is expected
to be more isotropic. Advantage is taken of these features through
the variable 
${\cal D}=[\Delta\phi(\mpt,\mathrm{jet_1}) + \Delta\phi(\mpt,\mathrm{jet_2})]/2$.
For signal events, as well as for the non-MJ 
backgrounds, 
${\cal D} > \pi/2$ in the vast majority of events,
whereas the MJ background events tend
to be symmetrically distributed around $\pi/2$.
In the analysis sample, ${\cal D} > \pi/2$ is therefore required.
To improve the efficiency of this criterion for the $(W\to\mu\nu)H$ signal 
with non-identified muons, 
tracks satisfying isolation criteria are removed from the $\mpt$ computation.
The reverse of the $\cal D$ requirement is also used to define the MJ-model sample,
as explained below.

Events containing an isolated electron or muon with $\pt > 15$~GeV are
rejected to ensure there is no overlap with the D0 $WH$ search in the
lepton$+\met$ topology~\cite{WHpub}.

The EW control sample is selected 
in a similar manner to the analysis sample, except that 
an isolated muon with $\pt>15$~GeV is required. 
The multijet content of this sample is rendered negligible by requiring 
that the transverse mass of the muon and \met\ system is larger 
than 30~GeV and that the $\met$,
calculated taking account of the muon from the $W$ boson decay, 
is greater than $20~$GeV.
To ensure similar jet topologies for the analysis and 
EW control samples, the \met, not corrected for the selected muon,  
is required 
to exceed 40~GeV. The number of selected events is in good agreement with the SM expectation. 
All the kinematic distributions are also well described once reweightings of the distributions 
of $\Delta\eta(\mathrm{jet_1},\mathrm{jet_2})$ and $\eta(\mathrm{jet_2})$
are performed, as suggested by a comparison~\cite{Zjets} of data with
a simulation of $(W/Z)$+jets using the {\sc sherpa} generator~\cite{sherpa}.
The distribution of the dijet mass in the EW control sample is shown in 
Fig.\ \ref{JetDR_ew_pretag}(a).

\mathchardef\mhyphen="2D

The MJ-model sample, used to determine the MJ background,
is selected in the same manner as the analysis sample, 
except that the requirement ${\cal D}>\pi/2$ is reversed.
The small remaining contributions from non-MJ SM background processes 
in the ${\cal D}<\pi/2$ region are subtracted, and the resulting sample 
is used to model the MJ background in the analysis sample. 
The MJ background in the region ${\cal D}>\pi/2$ is normalized by performing a fit of the 
sum of the MJ and SM backgrounds to the \met\ distribution of the data in the analysis sample. 
 
The MJ-enriched sample is used to test the validity of this approach and 
is defined in the same manner as the analysis sample, except that 
${\cal S}<4.5$ is now required (see Fig.\ \ref{HM_analysis_pretag}). 
As a result, the MJ background dominates the 
entire range of $\cal D$ values, and this sample is used to verify that 
the events with ${\cal D}<\pi/2$ correctly model those with ${\cal D}>\pi/2$. 
The distribution of the dijet mass in the MJ-enriched sample is shown in
Fig.\ \ref{JetDR_ew_pretag}(b).

\begin{figure}[tp]
\includegraphics[width=8.5cm]{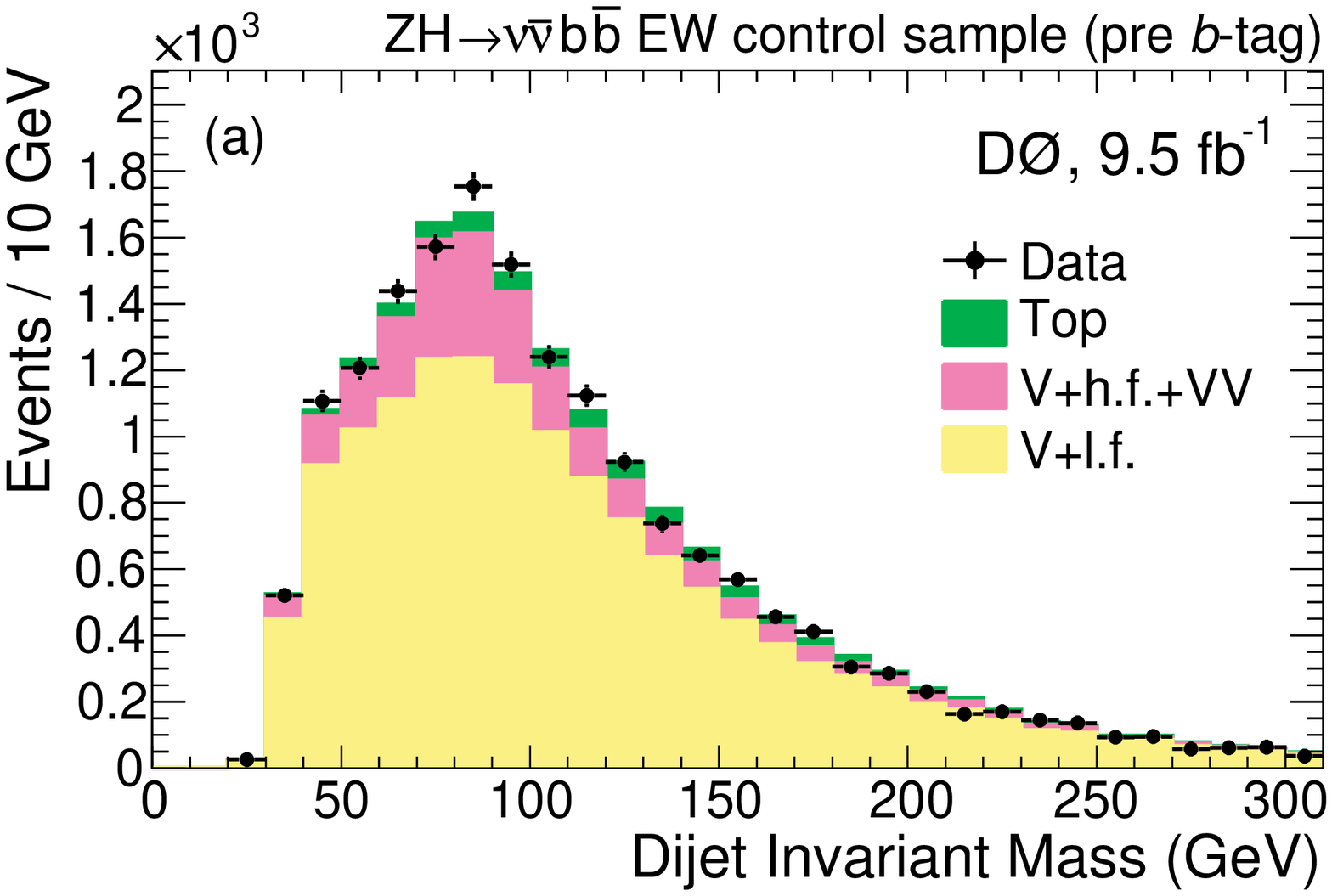}
\includegraphics[width=8.5cm]{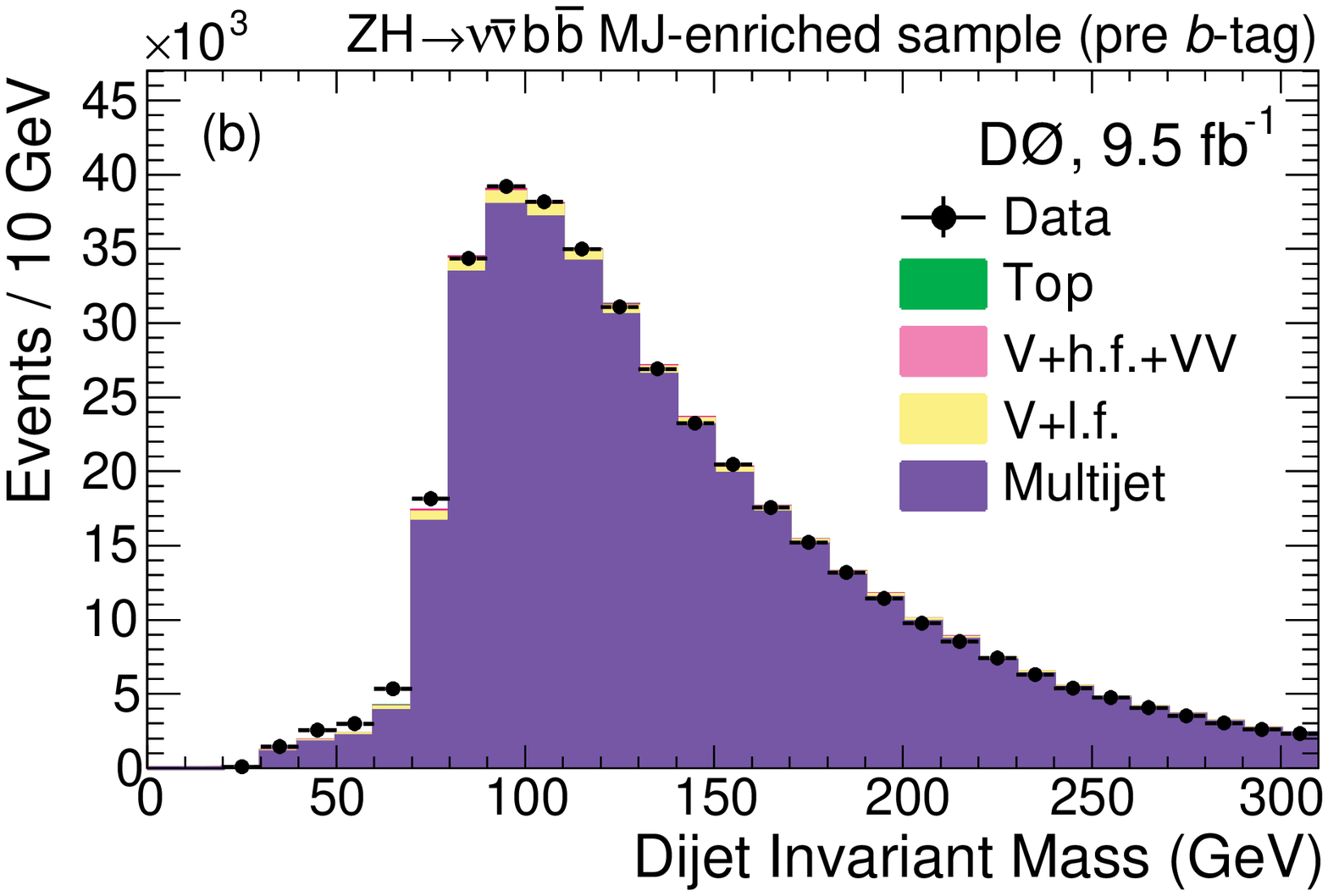}
\caption{\label{JetDR_ew_pretag}
(Color online.) Distributions of the dijet mass before $b$-tagging in the 
(a) EW control and (b) MJ-enriched samples. 
The data are shown as points with error bars and the background contributions as histograms: 
dibosons are labeled as ``VV,'' ``V+l.f.''\ includes $(W/Z)$+$(u,d,s,g)$ jets, 
``V+h.f.''\ includes $(W/Z)$+$(b,c)$ jets, and ``Top'' 
includes pair and single top quark production.}
\end{figure}

A multivariate $b$-tagging discriminant, with several boosted decision trees 
as inputs, is used to select events with one or
more $b$ quark candidates. This algorithm is an upgraded version of the
neural network $b$-tagging technique described in Ref.\ \cite{btag}. The new
algorithm includes more information related to the lifetime of the
jet and results in a better discrimination between $b$ and light ($u,d,s,g$)
jets. It provides an output between 0 and 1 for each
jet, with a value closer to one indicating a higher probability that
the jet originated from a $b$ quark. The output from the algorithm
measured in simulated events is adjusted to match the output measured 
in dedicated data samples as described in more detail in Ref.\ \cite{btag}. 
>From this continuous output, thirteen operating points ($L_b = 0, 1, \ldots ,12$) 
are defined, with $b$ purity increasing with $L_b$. Jets with $L_b=0$ are defined as untagged.
The typical per-$b$-jet efficiency and misidentification rate for light-flavor jets
are about 80\% (50\%) and 10\% (1\%) for the loosest non-zero (tightest) 
$b$-tag operating point, respectively.

To improve the sensitivity of the analysis, two high signal purity samples are 
defined from the analysis sample 
using the variable $L_{bb}=L_b(\mathrm{jet_1})+L_b(\mathrm{jet_2})$. 
The two samples are defined as follows:
a tight $b$-tag sample with $L_{bb} \geq 18$ and
a medium $b$-tag sample with $ 11 \leq L_{bb} \leq 17$.
The medium $b$-tag sample contains events with two loosely $b$-tagged jets, 
as well as events with one tightly $b$-tagged jet and one untagged jet.
The signal-to-background ratios for a Higgs boson mass of 125~GeV in the pre, 
medium, and tight $b$-tag
samples, after applying a multijet veto (defined in the next section),
are respectively $0.035\%$, $0.23\%$, and $1.00\%$.

\section{Analysis using decision trees}\label{sec:dt}

A stochastic gradient boosted decision tree (DT) technique is employed,
as implemented in the {\sc tmva} package~\cite{TMVA},
to improve the discrimination between signal and background processes.

First, an ``MJ DT'' (multijet-rejection DT) is trained to
discriminate between signal and MJ-model events before $b$-tagging
is applied. To avoid Higgs boson mass dependence at this stage
of the analysis, signal events are not used, and the MJ DT is trained
on a sample of $(W/Z)$+heavy-flavor jets events instead. 
Variables that provide some discrimination have been
chosen for the MJ DT, excluding those strongly correlated to the Higgs
mass (such as the dijet mass itself or the 
$\Delta R=\sqrt{(\Delta \eta)^2+(\Delta \phi)^2}$ between $\mathrm{jet_1}$ and $\mathrm{jet_2}$). 

\begin{figure}[htp]
\includegraphics[width=8.5cm]{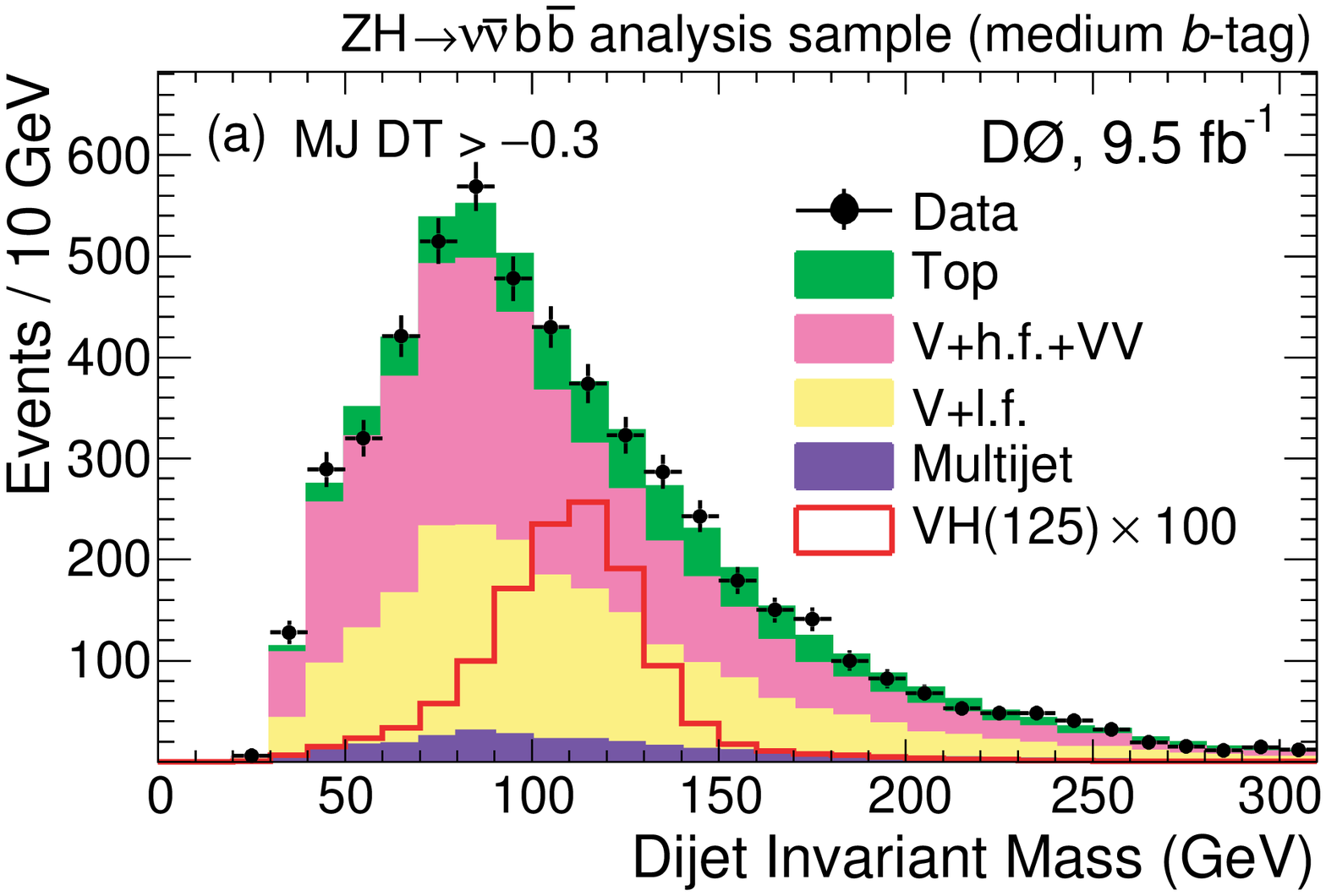}
\includegraphics[width=8.5cm]{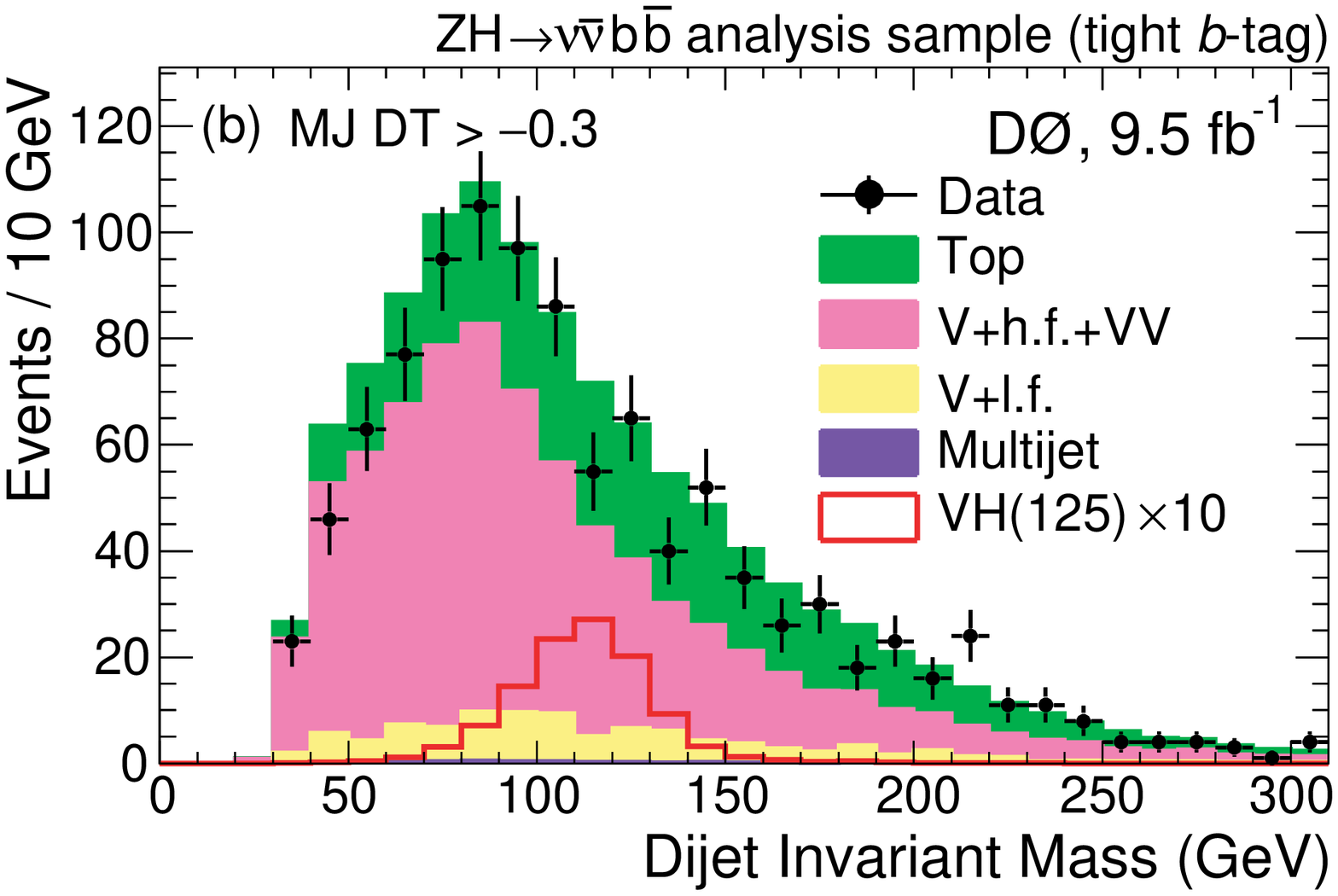}
\caption{\label{DIM_2tag}
(Color online.) Dijet invariant mass in the analysis sample after the multijet veto for events with 
(a) medium $b$-tag and (b) tight $b$-tag. The distributions for signal (VH), which are
scaled by a factor of 100 for medium $b$-tag and 10 for tight $b$-tag respectively, include
$ZH$ and $WH$ production for $M_H=125$~GeV. The data are shown as points with error bars
and the background contributions as histograms: dibosons are labeled as ``VV,'' ``V+l.f.''\ includes $(W/Z)$+$(u,d,s,g)$ jets, ``V+h.f.''\ includes $(W/Z)$+$(b,c)$ jets, and ``Top'' includes pair and single top quark production.}
\end{figure}

The MJ DT output, which ranges between $-1$ and $+1$, is shown in Fig.\ \ref{MJdecision} for the
analysis sample
after the medium $b$-tagging requirement. 
Good agreement is seen between data and the
predicted background.
A value of the multijet
discriminant in excess of $-0.3$ is required (multijet veto), which
removes 93\% of the multijet background while retaining 
90\% of the signal for $M_H=125$~GeV.
The numbers of expected signal and background events, as well as the number of
observed events, are given in Table~\ref{yields} after imposing the
multijet veto. Dijet mass distributions in the analysis sample after the multijet
veto are shown 
in Fig.\ \ref{DIM_2tag} for $b$-tagged events.

\begin{figure}[htp]
\includegraphics[width=8.5cm]{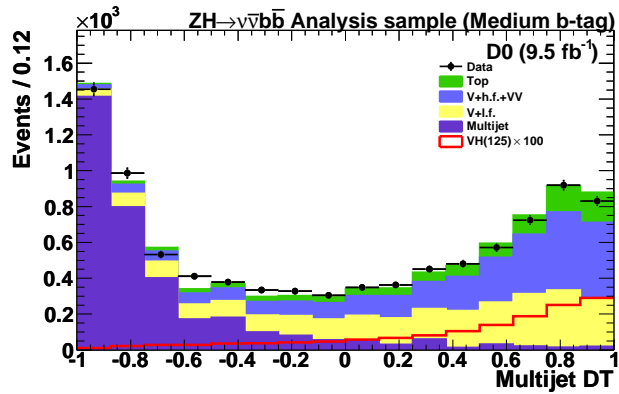}
\caption{\label{MJdecision}
(Color online.) Distribution of the MJ DT output after the medium $b$-tagging requirement in the analysis sample. 
The distribution for signal (VH),  shown for $M_H=125$~GeV, is scaled by a 
factor of 100 and includes $ZH$ and $WH$ production.
The data are shown as points with error bars and the background contributions as histograms: 
dibosons are labeled as ``VV,'' ``V+l.f.''\ includes $(W/Z)$+$(u,d,s,g)$ jets, 
``V+h.f.''\ includes $(W/Z)$+$(b,c)$ jets, and ``Top'' includes pair and single 
top quark production.}
\end{figure}

\begin{table*}
\caption{\label{yields} The numbers of expected signal, expected background, and observed 
data events after the 
multijet veto, for the pre, medium, and tight $b$-tag samples. 
The signal corresponds to $M_H=125$~GeV, ``Top'' includes pair and 
single top quark production, and ``$VV$'' is the sum of all diboson
processes. The uncertainties quoted on the signal and total background 
arise from the statistics of the simulation 
and from the sources of systematic uncertainties mentioned in the text.}
\begin{tabular}{lr@{$\,\pm \,$}llr@{$\,\pm \,$}llcccccr@{$\,\pm \,$}llc}
\hline\hline
Sample & \multicolumn{3}{c}{$ZH$}   &  \multicolumn{3}{c}{$WH$}   & $W+$jets & $Z+$jets &  Top &  $VV$ &  MJ &  \multicolumn{3}{c}{Total Background}   & Observed \\
\hline
Pre $b$-tag & 18.3&1.8  & & 16.7&1.6 & & 66895 & 25585 & 1934 & 3144 & 1977 & 99535&12542 & & 98980\\
Medium $b$-tag & 6.7&0.7 & & 6.1&0.6 & & 3112 & 1074 & 761 & 237 & 278 & 5462&776 & & 5453\\
Tight $b$-tag & 6.0&0.8 & & 5.3&0.7 & & 443 & 252 & 377 & 56 & 6 & 1134&192 & & 1039\\
\hline\hline
\end{tabular}
\end{table*}

\begin{figure}[b]
\includegraphics[width=8.5cm]{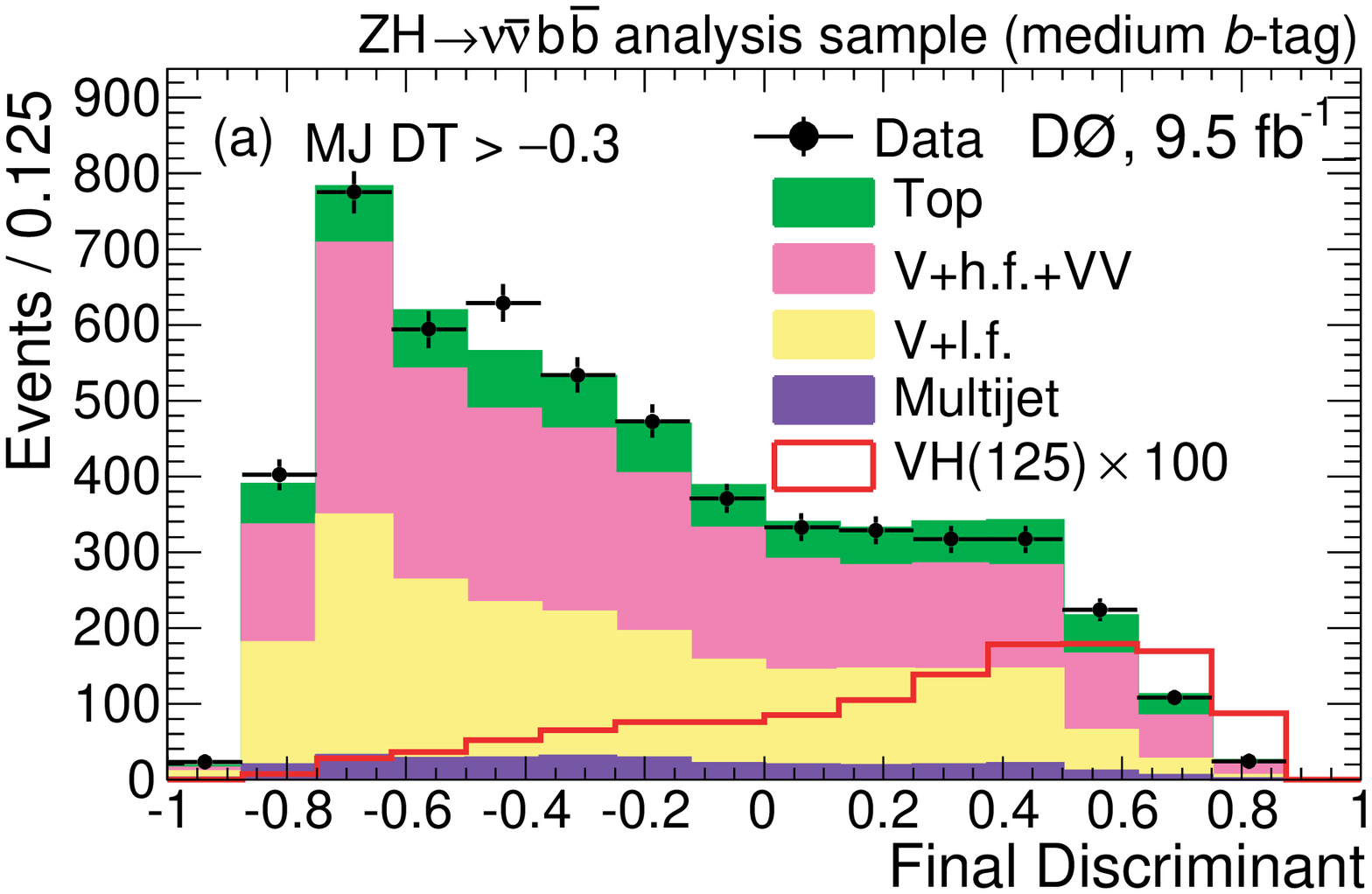}
\includegraphics[width=8.5cm]{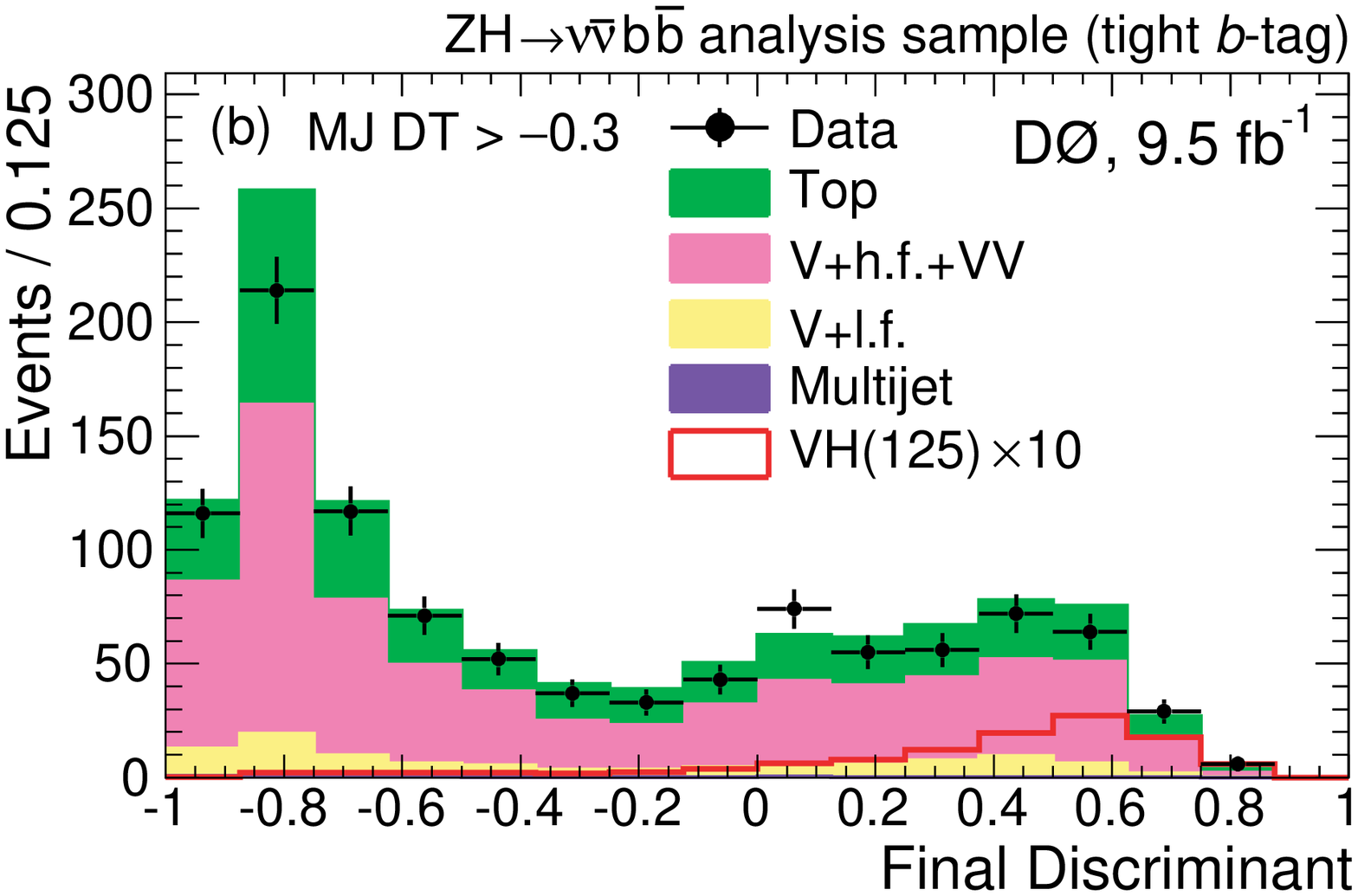}
\caption{\label{decision}
(Color online.) The SM DT output for the $(W/Z)H$ search with $M_H=125$~GeV following the
multijet veto for events with (a) medium $b$-tag and (b) tight $b$-tag prior to the fit to data. 
The distributions for signal (VH) are scaled by a
factor of 100 for medium $b$-tag events and 10 for tight $b$-tag events, respectively,
and include $ZH$ and $WH$ production for $M_H=125$~GeV. The data are shown as points
with error bars 
and the background contributions as histograms: dibosons are labeled as ``VV,'' 
``V+l.f.''\ includes $(W/Z)$+$(u,d,s,g)$ jets,
``V+h.f.''\ includes $(W/Z)$+$(b,c)$ jets, and ``Top'' 
includes pair and single top quark production.}
\end{figure}

Next, to separate signal from the remaining SM backgrounds, two ``SM
DTs'' (SM-background-rejection DTs) are trained for each $M_H$, one in the
medium $b$-tag channel and one in the tight $b$-tag channel. Some of
the MJ DT input variables are used again, but most of the
discrimination comes from additional kinematic variables 
correlated to
the Higgs boson mass, of which, as expected, the dijet mass has the strongest 
discriminating power. 
The SM DT outputs, 
which range between $-1$ and $+1$, are used as final discriminants. 
Their distributions are shown in Fig.\ \ref{decision} for $M_H=125$~GeV.

\section{Systematic uncertainties} \label{sec:systs}

Experimental uncertainties arise from the integrated luminosity
(6\%)~\cite{lumi}, the trigger simulation (2\%), the jet energy calibration 
and resolution [(1--2)\%], 
jet reconstruction and taggability (3\%), the lepton identification
(1\%), the modeling of the MJ background (25\%, which translates into
a 1\% uncertainty on the total background), and the $b$-tagging 
(from 4\% for background in the medium $b$-tag sample
to 9\% for signal in the tight $b$-tag sample).
In addition to the impact of these uncertainties on the integrated signal and background 
yields mentioned above,
modifications of the shapes of the final discriminants are also considered, when relevant.
Correlations among systematic uncertainties in signal and each background are taken into 
account when extracting the final results.

Theoretical uncertainties on cross sections for SM processes are 
estimated as follows.
For $(W/Z)$+jets production, an uncertainty of 10\% 
is assigned to the total cross sections and an uncertainty 
of 20\% to the heavy-flavor fractions (estimated using {\sc mcfm} at NLO~\cite{mcfm}). 
For other SM backgrounds, uncertainties are taken from Ref.~\cite{xsections} 
or using {\sc mcfm}~\cite{mcfmdiboson} and range from 6\% to 10\%. 
The uncertainties on cross sections for signal (7\%)
are taken from Ref.~\cite{signal}. Uncertainties on the shapes of the final 
discriminants arise from (i) the modeling of
$(W/Z)$+jets, assessed by varying the 
renormalization and factorization scales and by comparing results from {\sc alpgen} 
interfaced with {\sc herwig}~\cite{herwig} to {\sc alpgen}
interfaced with {\sc pythia}, and (ii) the
choice of PDFs, estimated using the prescription of Ref.~\cite{cteq}.

\section{Limit setting procedure} \label{sec:limset}

Agreement is found between data and the predicted background,
both in the numbers of selected events (Table~\ref{yields})
and in the distributions of final discriminants (Fig.\ \ref{decision}), 
once systematic uncertainties are taken into
account.
The modified frequentist CL$_{\mathrm s}$ approach~\cite{cls} is used to set limits on the
cross section for SM Higgs boson production, where the test statistic
is a log-likelihood ratio (LLR) for the background-only and
signal+background hypotheses. The result is obtained by summing LLR values over the
bins in the final discriminants shown in Fig.\ \ref{decision}.  The
impact of systematic uncertainties on the sensitivity of the analysis
is reduced by maximizing a ``profile'' likelihood function~\cite{wade}
in which these uncertainties are given Gaussian constraints associated
with their priors.
Figure~\ref{bkgdsubtract} shows a comparison of the
SM DT distributions expected for a signal with $M_H=125$ GeV
and observed for the background-subtracted data.
The subtracted background and its uncertainties are the result 
of the profile likelihood fit to the data 
under the background-only hypothesis. 

\begin{figure}[htp]
\includegraphics[width=8.5cm]{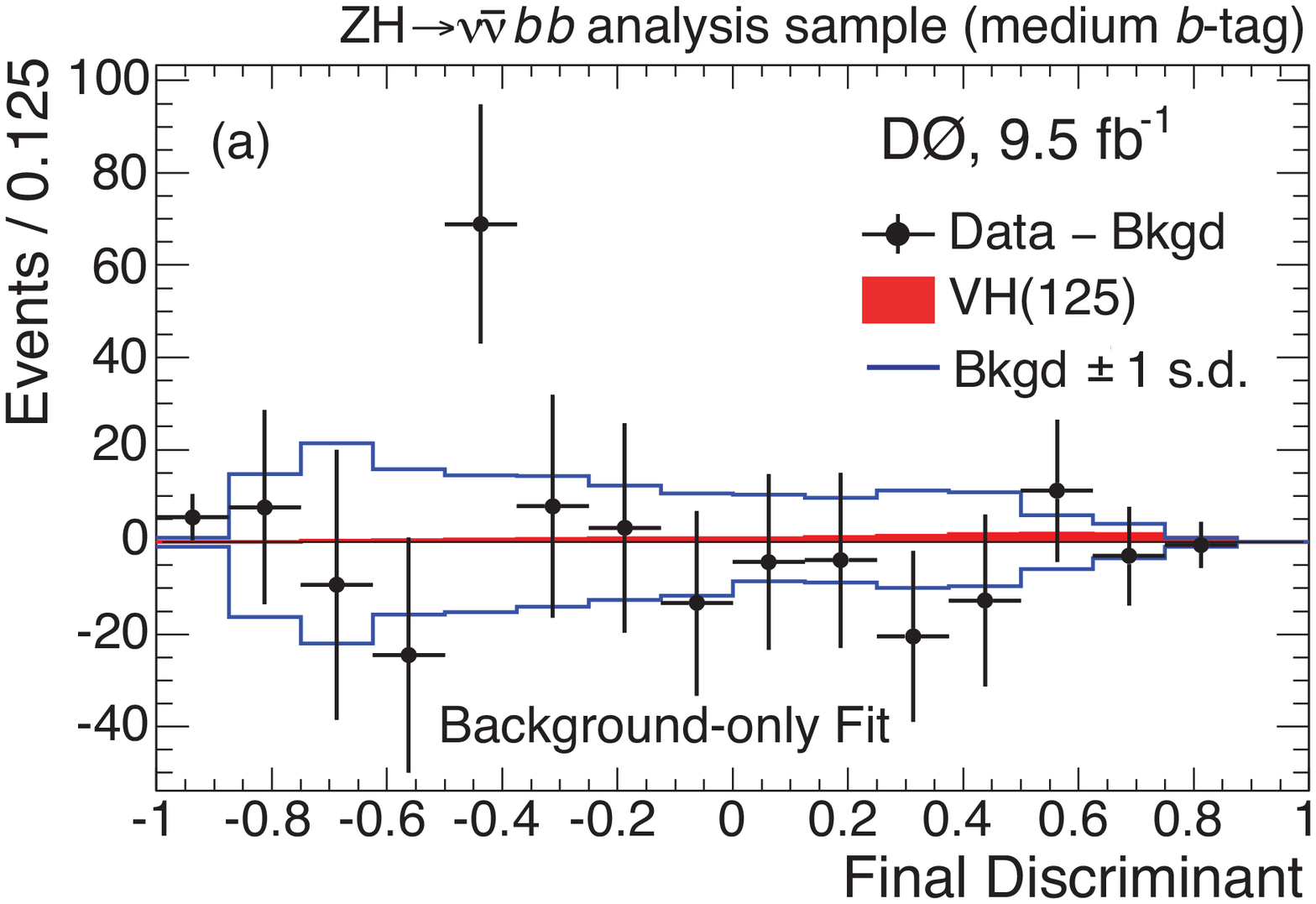}
\includegraphics[width=8.5cm]{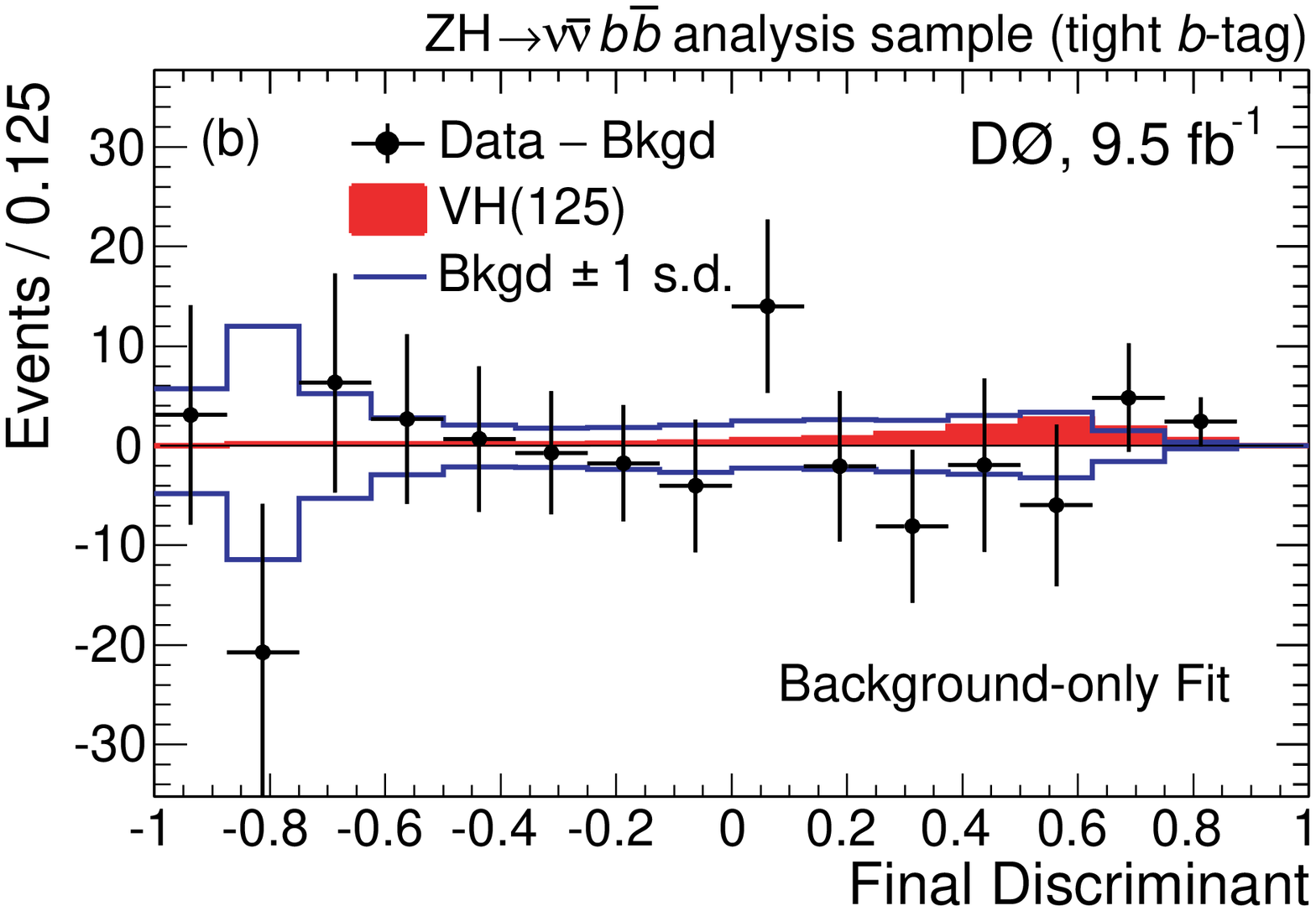}
\caption{\label{bkgdsubtract}
(Color online.) Final discriminant distributions expected for a SM VH signal 
with $M_H=125$~GeV (filled histogram)
and observed for background-subtracted data (points with statistical error bars) 
for the (a) medium and (b) tight $b$-tag channels.
The subtracted background is the result of the profile likelihood fit to the data 
under the background-only hypothesis. Also shown is
the $\pm 1$ standard deviation (s.d.) band on the fitted background.
No scaling factor is applied to the signal.
}
\end{figure}

\section{Higgs Boson Search Results} \label{sec:higres}

\begin{figure}[htp]
\includegraphics[width=8.5cm]{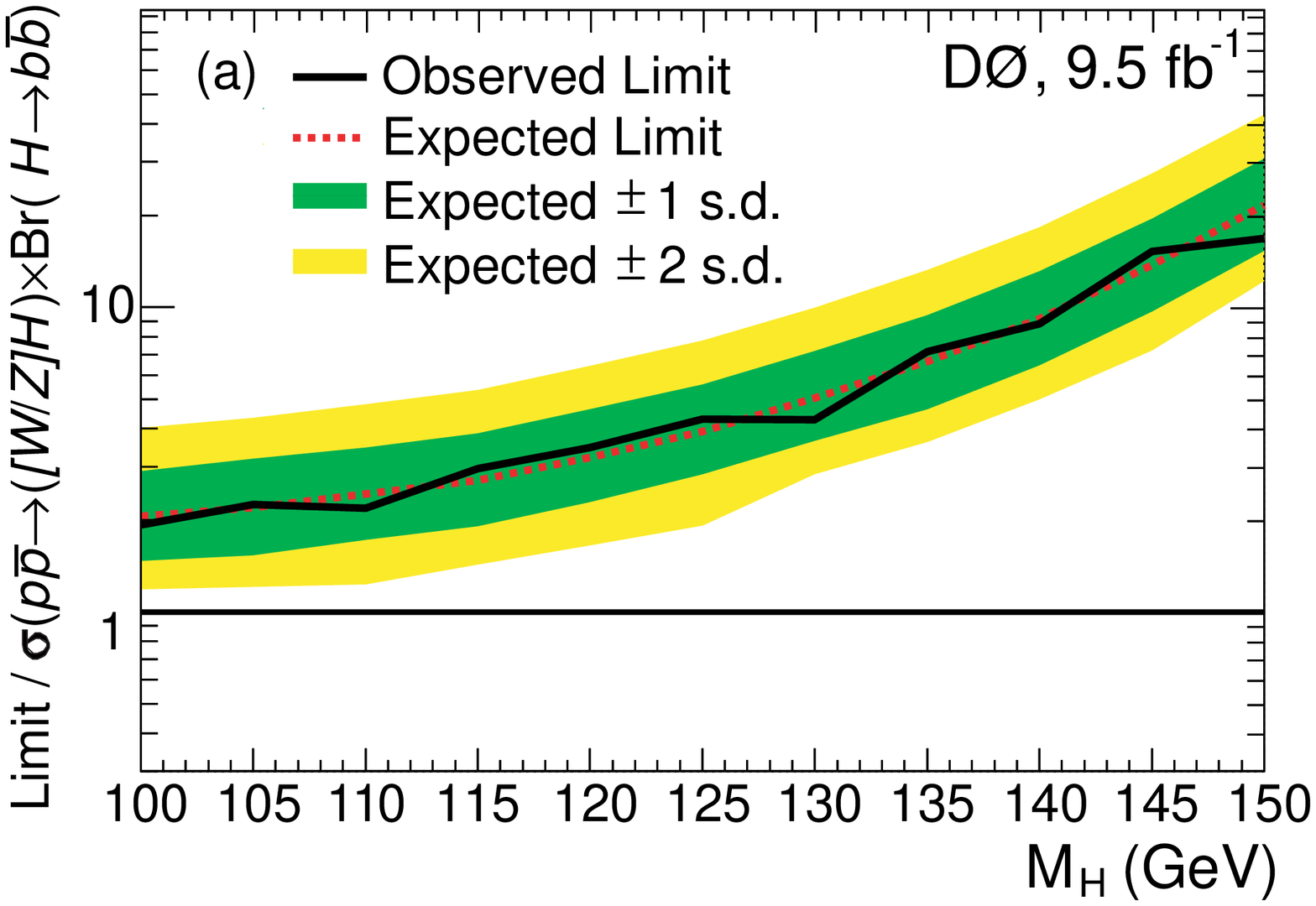}
\includegraphics[width=8.5cm]{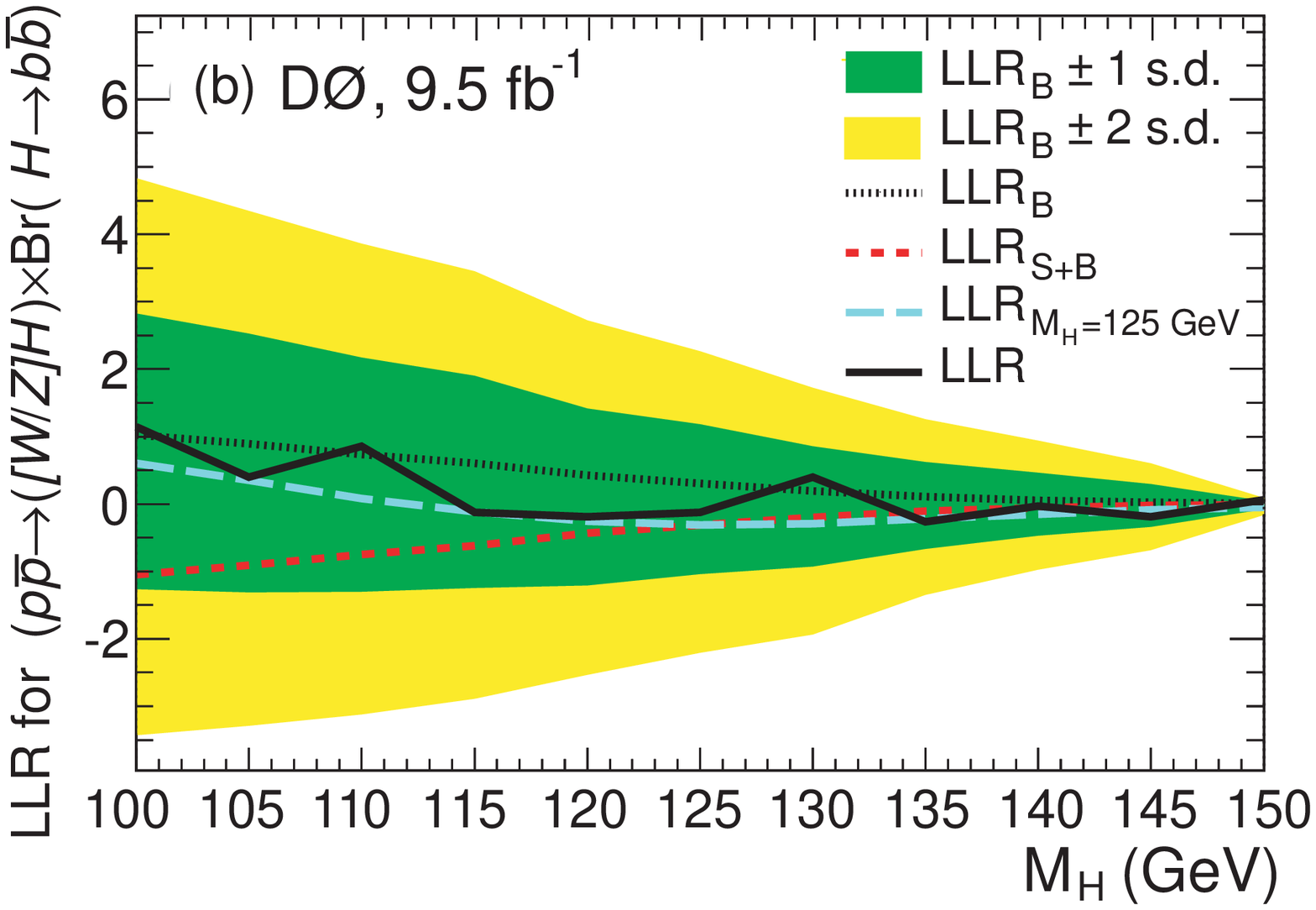}
\caption{\label{final} 
(Color online.) (a) Ratio of the observed (solid black) and expected (dotted red) 
exclusion limits to the SM production cross section. 
(b) The observed (solid black) and expected 
LLRs for the background-only (black dots) and 
signal+background hypotheses (short red dashes), 
as well as the LLR expected in the presence of a Higgs boson with $M_H=125$~GeV (long blue dashes).
All are shown as a function of the tested value of $M_H$ with 
the green and yellow shaded areas 
corresponding to the 1 and 2 standard deviation (s.d.) variations around the background-only hypothesis.}
\end{figure}

The results are given as limits in
Table~\ref{limits_runIIb} and Fig.~\ref{final}(a) and
in terms of LLR values in Fig.\ \ref{final}(b). For $M_H=125$~GeV,
the observed and expected limits on the combined cross section of $ZH$
and $WH$ production are factors of 4.3 and 3.9 larger than the SM
value, respectively, assuming SM branching fractions.
In Fig.\ \ref{final}(b), the median expected LLR in the presence of a Higgs boson with a mass of 125~GeV is also shown for comparison.

\begin{table*}
\caption{\label{limits_runIIb} 
The expected and observed upper limits measured using 9.5~\invfb\ of 
integrated luminosity on the $ZH$ plus $WH$
production cross section relative to the SM expectation,
assuming SM branching fractions, 
as a function of $M_H$.}
\begin{ruledtabular}
\begin{tabular}{lccccccccccc}
$m_{H}$ (GeV)  &100 &105 &110 &115 &120 &125 &130 &135 &140 &145 &150 \\
\hline
Expected &
2.1  &
2.2  &
2.4  &
2.7  &
3.2  &
3.9  &
5.0  &
6.7  &
9.2  &
13.8  &
21.6  \\
Observed &
1.9  &
2.3  &
2.2  &
3.0  &
3.5  &
4.3  &
4.3  &
7.2  &
8.8  &
15.3  &
16.8  \\
\end{tabular}
\end{ruledtabular}
\end{table*}

\section{Diboson Search Results} \label{sec:dibres}

The final states arising from the SM production of 
$(Z\to\nunub)(Z\to\bbb)$ and $(W\to\ell\nu)(Z\to\bbb)$ 
are the same in particle content and topology as 
those used for the Higgs boson search reported above when the lepton from $W\to\ell\nu$ is not reconstructed. 
Evidence for 
$ZZ$ and $WZ$ production can therefore be used 
to validate the techniques employed in the Higgs boson search. The 
only modification to the analysis is in the 
training of the final discriminants, where $ZZ$ and $WZ$ are now 
treated as signal with the remaining diboson 
process, $WW$, kept as background. 
A cross section scale factor of $0.94 \pm 0.31\thinspace\mathrm{(stat)} \pm 0.34\thinspace\mathrm{(syst)}$ 
is measured with respect to the 
predicted SM value of $(4.4 \pm 0.3)$~pb~\cite{mcfmdiboson}, 
with an observed (expected) significance 
of 2.0 (2.1) standard deviations. 

The measurement of the diboson cross section has also been carried out using as final discriminants the 
distributions of dijet invariant mass (as opposed to the SM DTs) in the medium and tight $b$-tag samples. 
A cross section scale factor of $1.08 \pm 0.35\thinspace\mathrm{(stat)} \pm 0.39\thinspace\mathrm{(syst)}$ 
is measured with respect to the predicted SM value, with an observed (expected) 
significance of 2.0 (1.9) standard deviations. The expected significance is slightly lower than the one
expected with the multivariate analysis, in which additional discrimination is provided by variables such
as the angular separation between jets or the event centrality.

Figure~\ref{bkgdsubtract_diboson_sbfit} shows the
final discriminant distributions in the medium and tight $b$-tag channels, 
as well as the dijet mass distribution summed over the medium and 
tight $b$-tag channels, for the expected $WZ+ZZ$ signal
and for the background-subtracted data.
The subtracted backgrounds and their uncertainties are the results 
of the profile likelihood fits to the data 
under the signal+background hypothesis. 

\begin{figure}[htp]
\includegraphics[width=8.5cm]{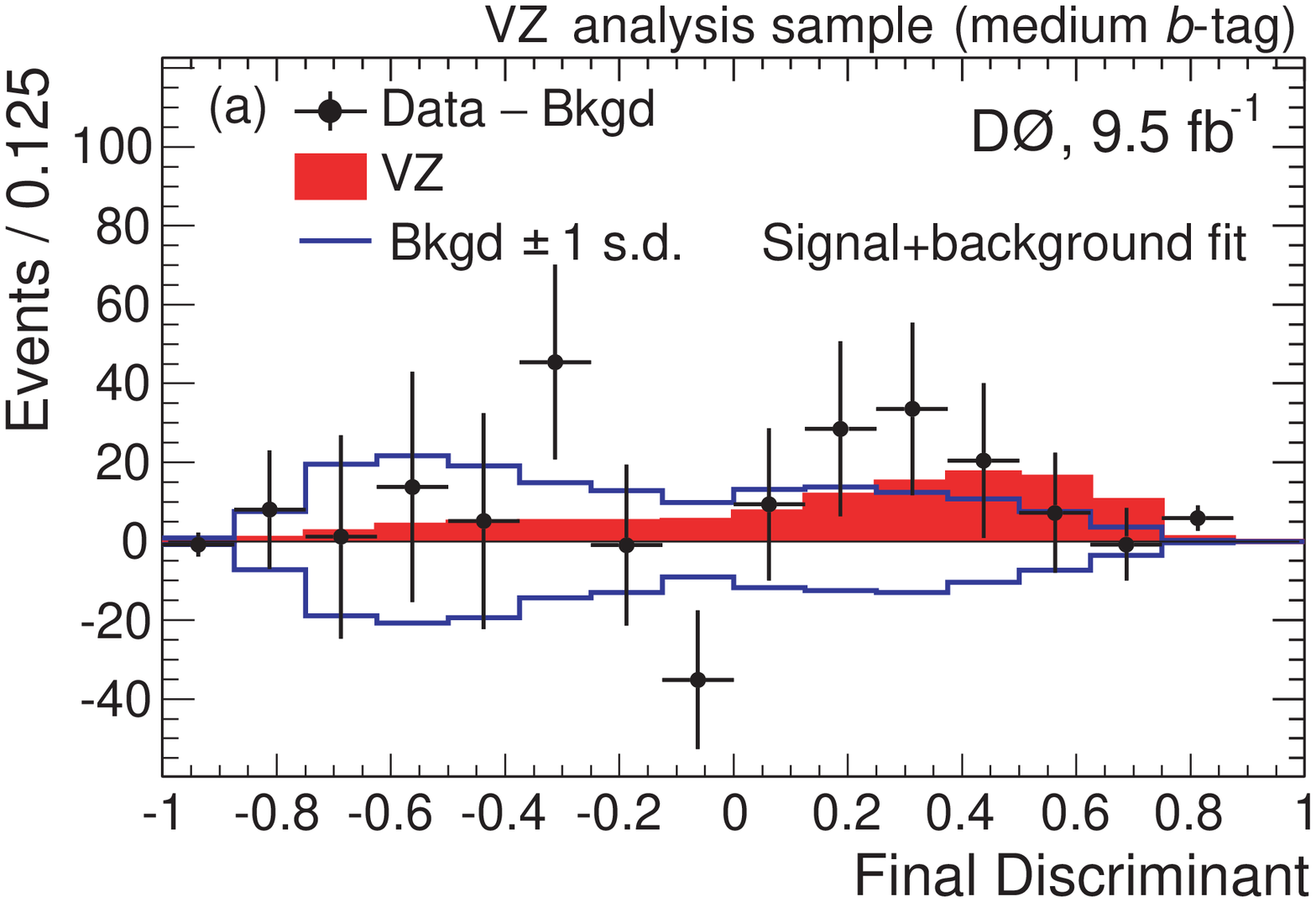}
\includegraphics[width=8.5cm]{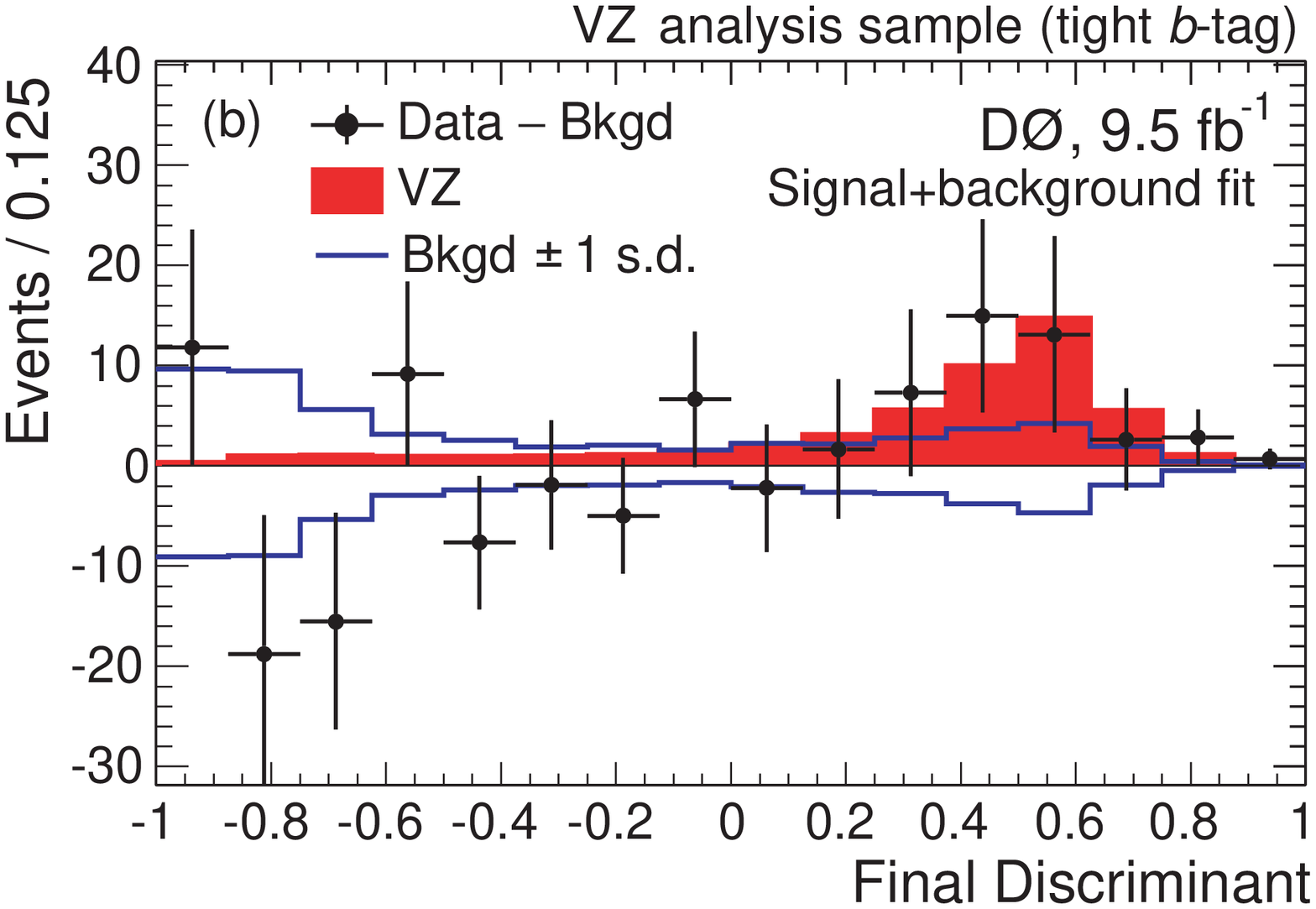}
\includegraphics[width=8.5cm]{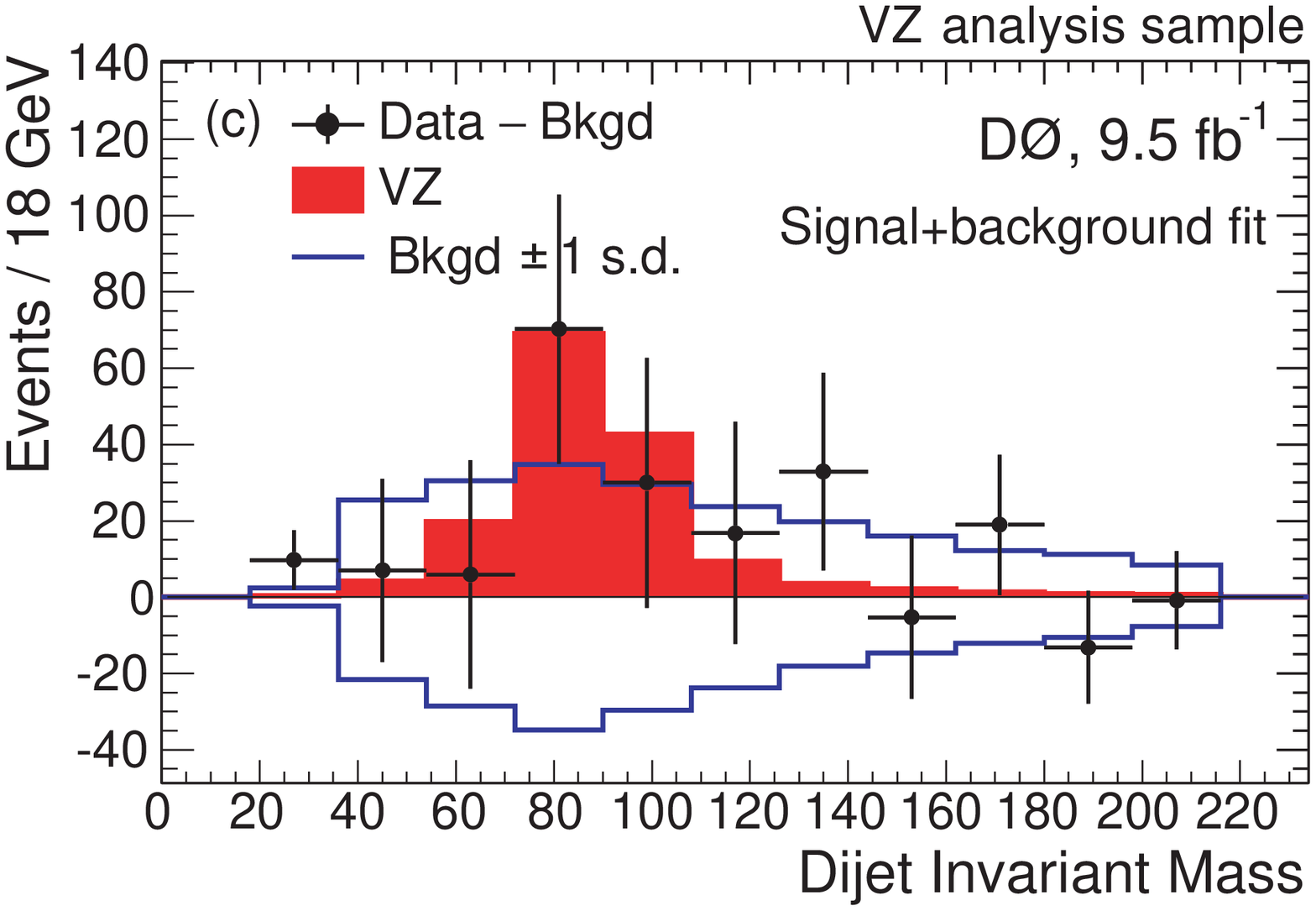}
\caption{\label{bkgdsubtract_diboson_sbfit}
(Color online.) Final discriminant distributions expected for a $WZ$ plus $ZZ$ signal (filled histogram) 
and observed for background-subtracted data (points with statistical error bars) 
for the (a) medium and (b) tight $b$-tag channels. 
(c) Similarly for the dijet mass distribution, 
summed over the medium and tight $b$-tag channels.
The subtracted backgrounds are the results of profile likelihood fits to the data 
under the signal+background hypothesis. Also shown are
the $\pm 1$ standard deviation (s.d.) bands on the fitted backgrounds.
The signal is scaled to the SM cross section.}
\end{figure}

\section{Summary}

We have performed a search for the standard model Higgs boson in
9.5~\invfb\ of \ppb\ collisions at $\sqrt{s}=1.96$~TeV collected with the 
D0 detector at the Fermilab Tevatron Collider. 
The final state considered contains a pair of $b$ jets and is characterized by an imbalance in
transverse energy, as expected from $\ppb\to ZH\to\nunub\bbb$ production and decays. 
The search is also 
sensitive to the $WH\to\ell\nu\bbb$ channel when the charged lepton is not 
identified. The data are found to be in good agreement with the expected background.
For a Higgs boson mass of 125~GeV, we set a limit at the 
95\% C.L. on the cross section $\sigma(\ppb\to [Z/W]H)$, assuming standard model
branching fractions, that is a factor of 4.3 larger than the
theoretical standard model value, for an expected factor of 3.9.

To validate our analysis techniques, we also performed a search for $WZ$ and $ZZ$ production, resulting in a measurement of the combined cross section that is a factor of 
$0.94 \pm 0.31\thinspace\mathrm{(stat)} \pm 0.34\thinspace\mathrm{(syst)}$ relative to the standard model prediction, with a significance of 2.0 standard deviations.

\section{Acknowledgements} \label{sec:acknow}
%
We thank the staffs at Fermilab and collaborating institutions,
and acknowledge support from the
DOE and NSF (USA);
CEA and CNRS/IN2P3 (France);
MON, NRC KI and RFBR (Russia);
CNPq, FAPERJ, FAPESP and FUNDUNESP (Brazil);
DAE and DST (India);
Colciencias (Colombia);
CONACyT (Mexico);
NRF (Korea);
FOM (The Netherlands);
STFC and the Royal Society (United Kingdom);
MSMT and GACR (Czech Republic);
BMBF and DFG (Germany);
SFI (Ireland);
The Swedish Research Council (Sweden);
and
CAS and CNSF (China).

\vspace*{0.5cm} 
 


\newpage





\appendix
\section{Supplementary material}

\subsection{Introduction}
No added material.


\subsection{Data and Simulated Samples}
No added material.


\subsection{Event selection}

The $\cal S$ distribution is shown for the analysis and EW-control samples in 
Fig.~\ref{HM_analysis_pretag}.

\begin{figure*}[htp]
\centering
\subfigure[]{\includegraphics[width=8.5cm]{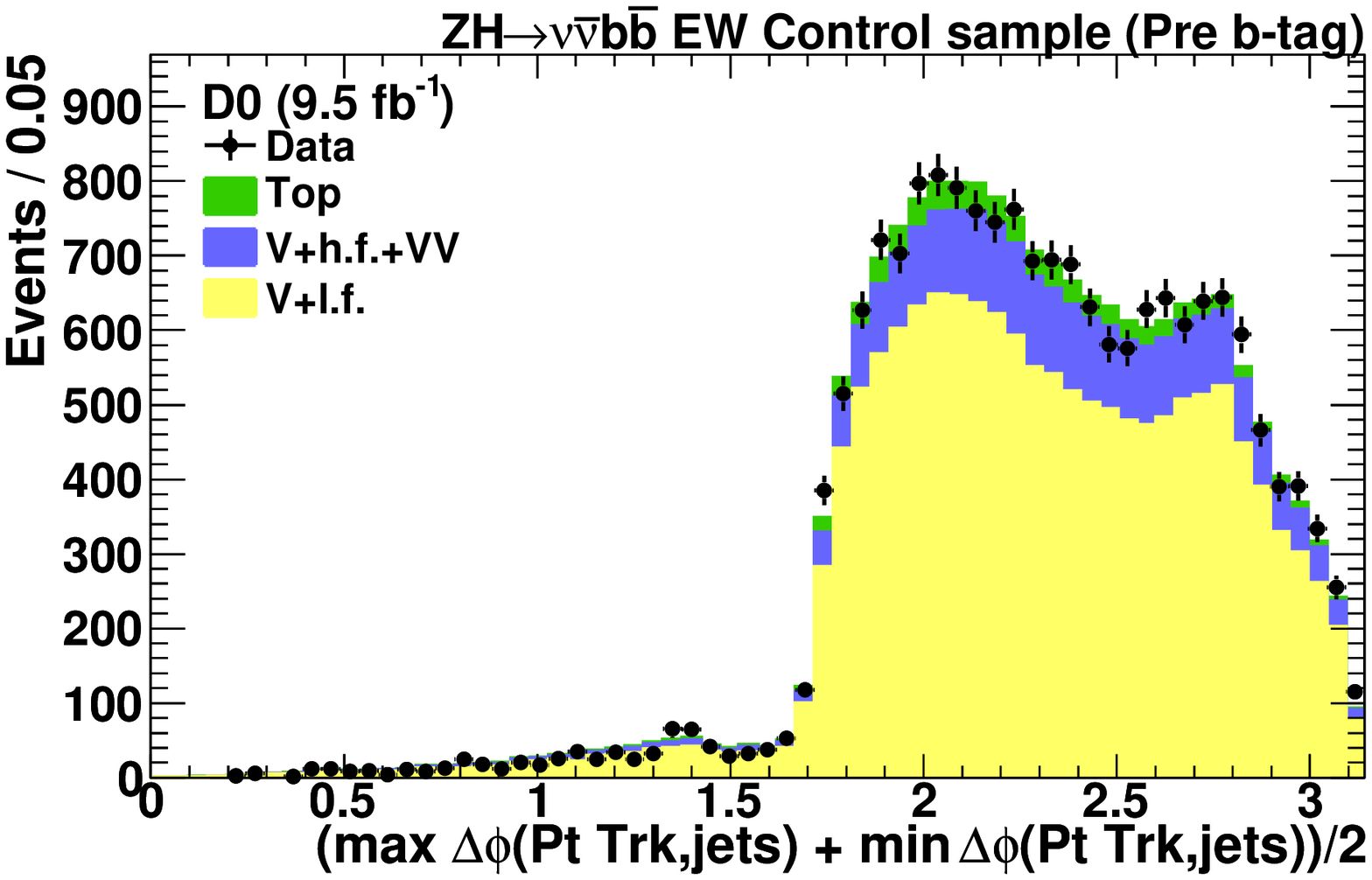}}
\subfigure[]{\includegraphics[width=8.5cm]{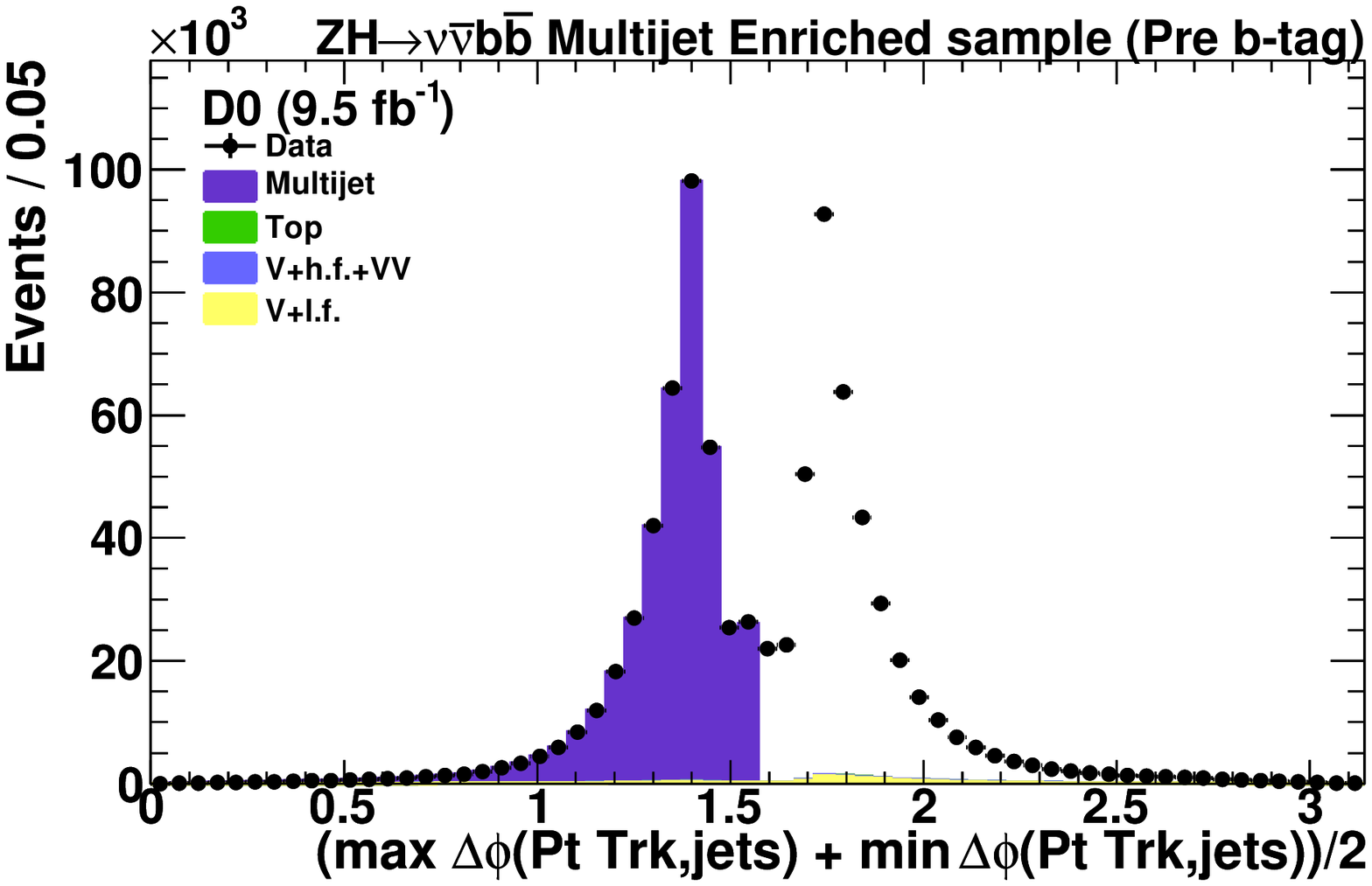}}
\caption{\label{Djet}  
Distribution of $\cal D$  in (a) the EW-control sample and (b) the MJ-enriched 
sample, without the requirement that it be larger than $\pi/2$. The data are 
shown as points and the background contributions as histograms: dibosons are 
labeled as ``VV,'' ``V+l.f.'' includes $(W/Z)$+$(u,d,s,g)$ jets, ``V+h.f.'' 
includes $(W/Z)$+$(b,c)$ jets and ``Top'' includes pair and single top quark 
production. In (b), the shaded region (${\cal D} < \pi/2$) is used to model 
the events in the unshaded region (${\cal D} > \pi/2$); the dip observed in 
the region around $\pi/2$ is due to the acoplanarity cut between the Higgs 
candidate jets. These distributions are shown before $b$ tagging.
}
\end{figure*}

\bigskip
The effectiveness of the use of the variable 
${\cal D}=(\Delta\phi(\mpt,\mathrm{jet_1}) + 
\Delta\phi(\mpt,\mathrm{jet_2}))/2$  
can be seen in Fig.~\ref{Djet}, where
the distribution of $\cal D$ is shown for the EW control
sample, dominated by events with real \met, and for the MJ-enriched sample,
dominated by events with \met\ arising from instrumental effects.
For signal events, as well as for the non-MJ 
backgrounds, it is expected that ${\cal D} > \pi/2$ in the vast 
majority of events, whereas the MJ background events tend
to be symmetrically distributed around $\pi/2$.
In the analysis sample, ${\cal D} > \pi/2$ is therefore required.

\begin{figure*}[htp]
\centering
\subfigure[]{\includegraphics[width=8.5cm]{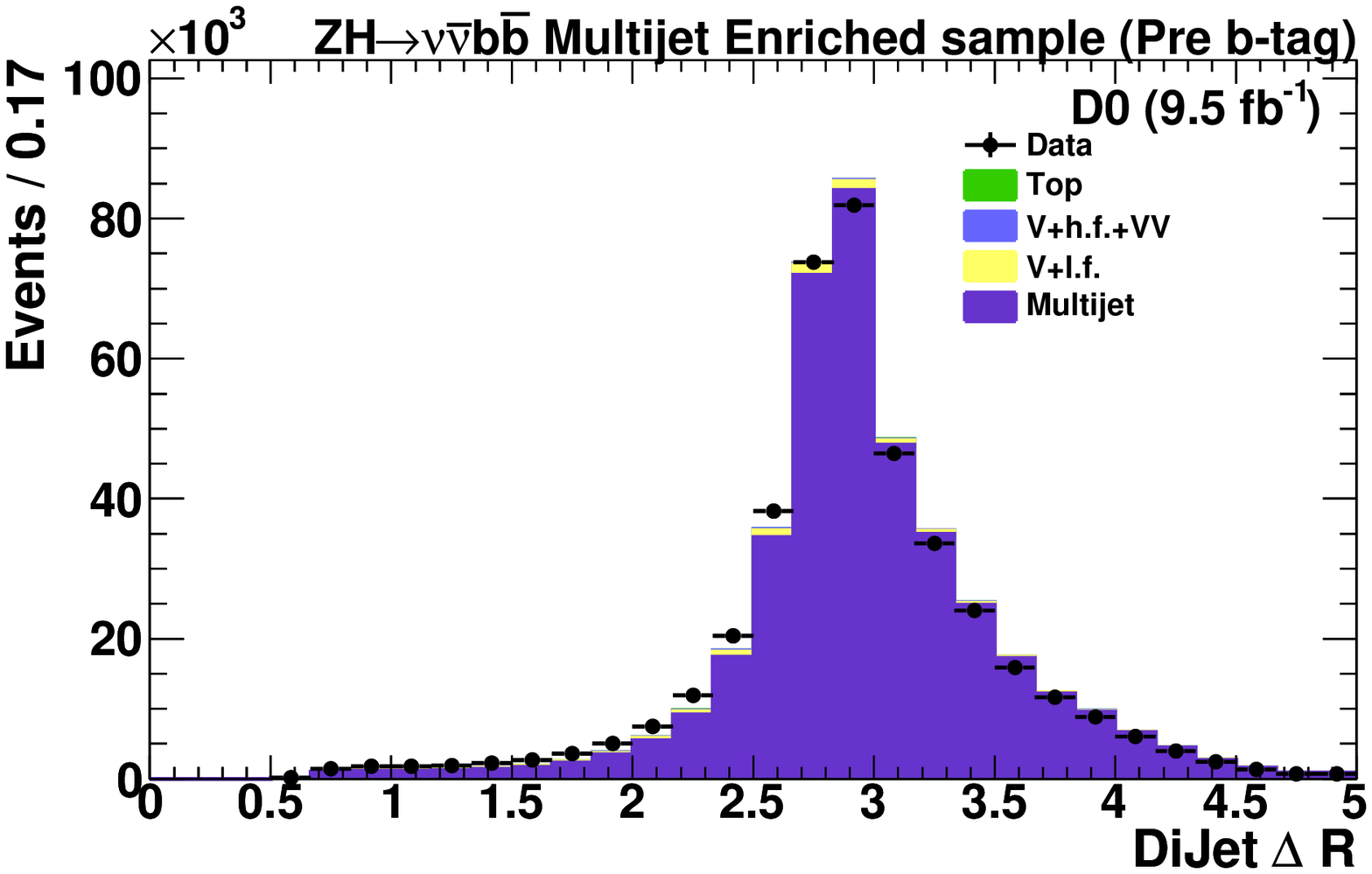}}
\subfigure[]{\includegraphics[width=8.5cm]{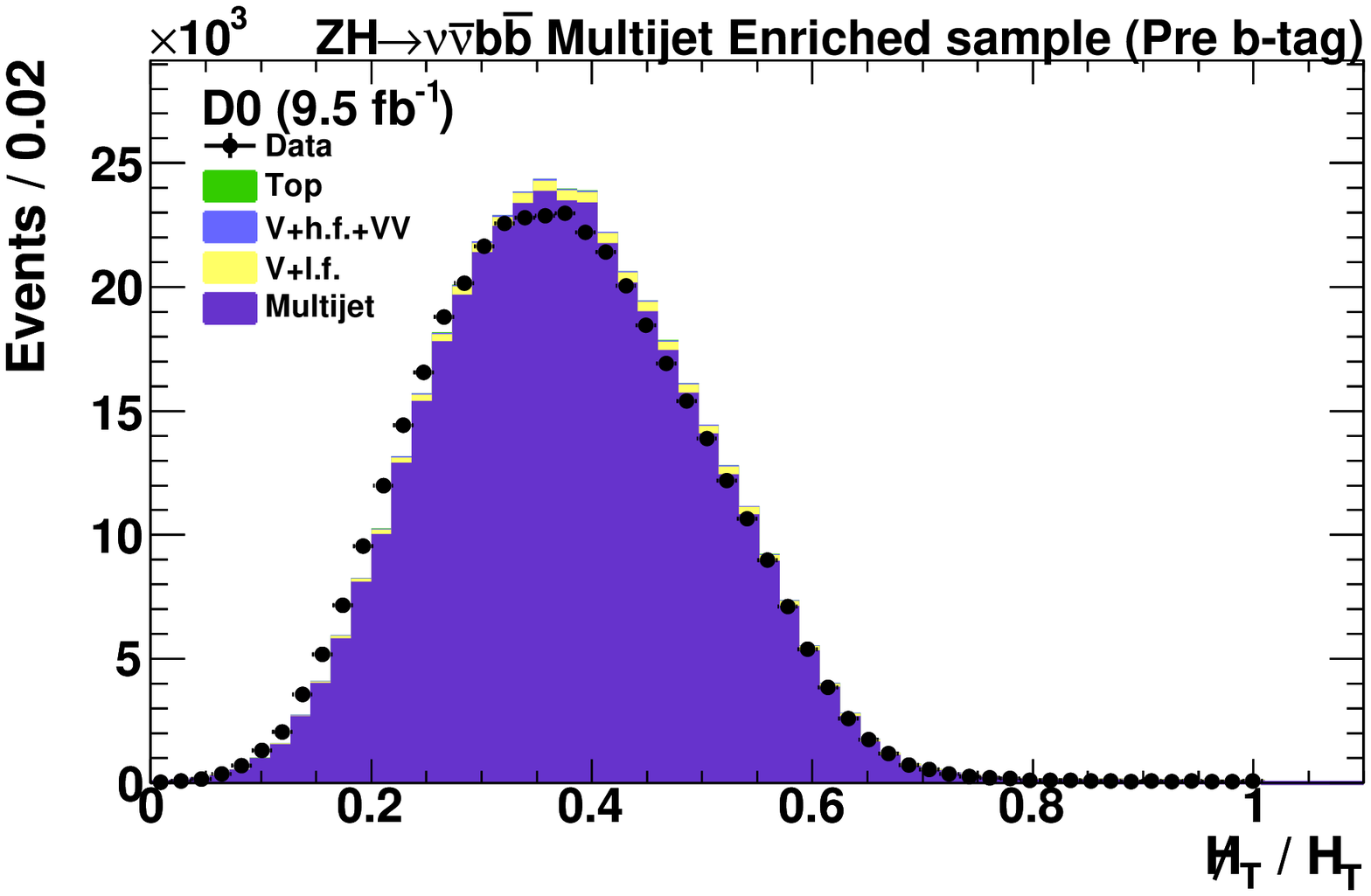}}
\subfigure[]{\includegraphics[width=8.5cm]{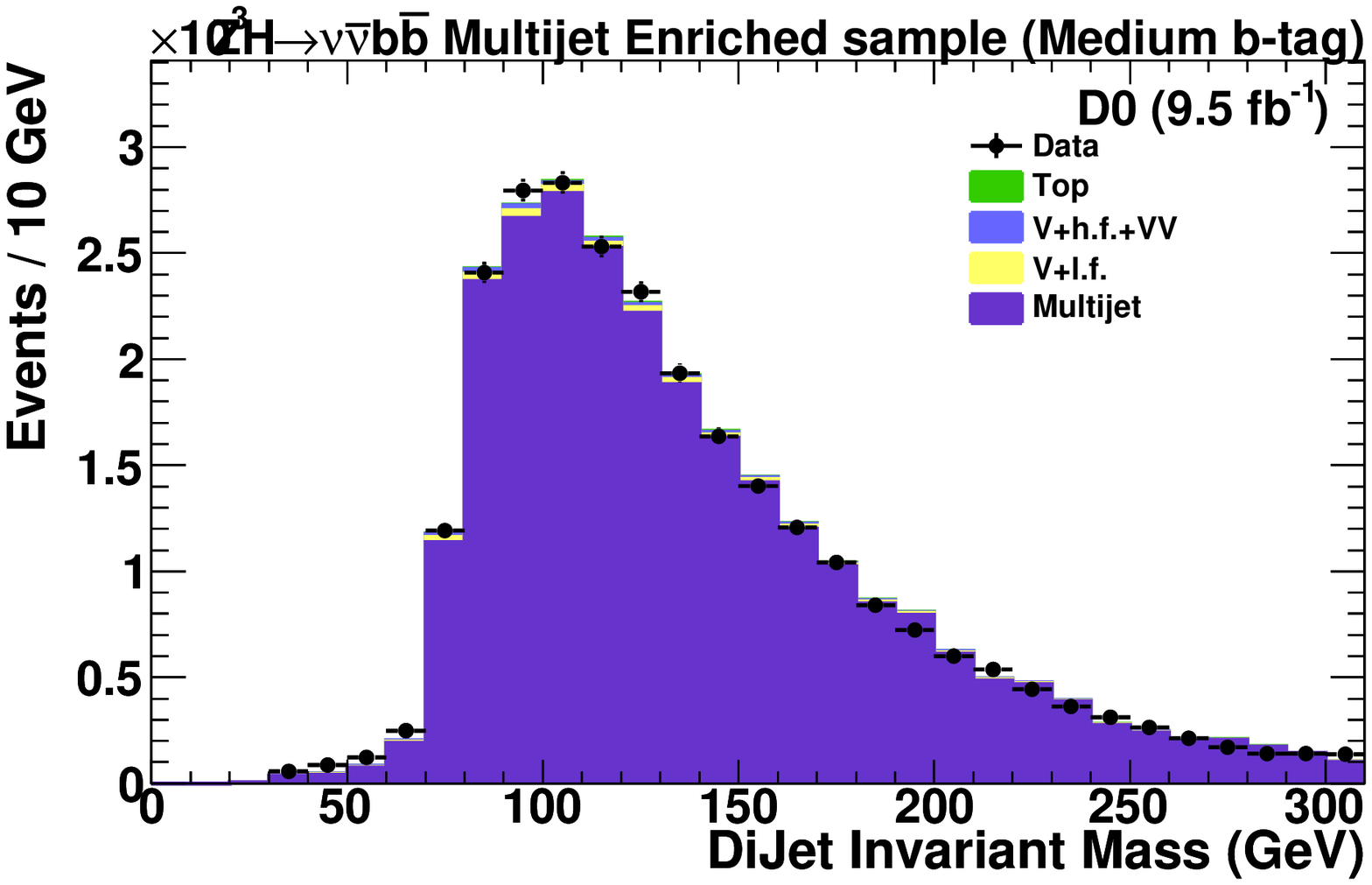}}
\subfigure[]{\includegraphics[width=8.5cm]{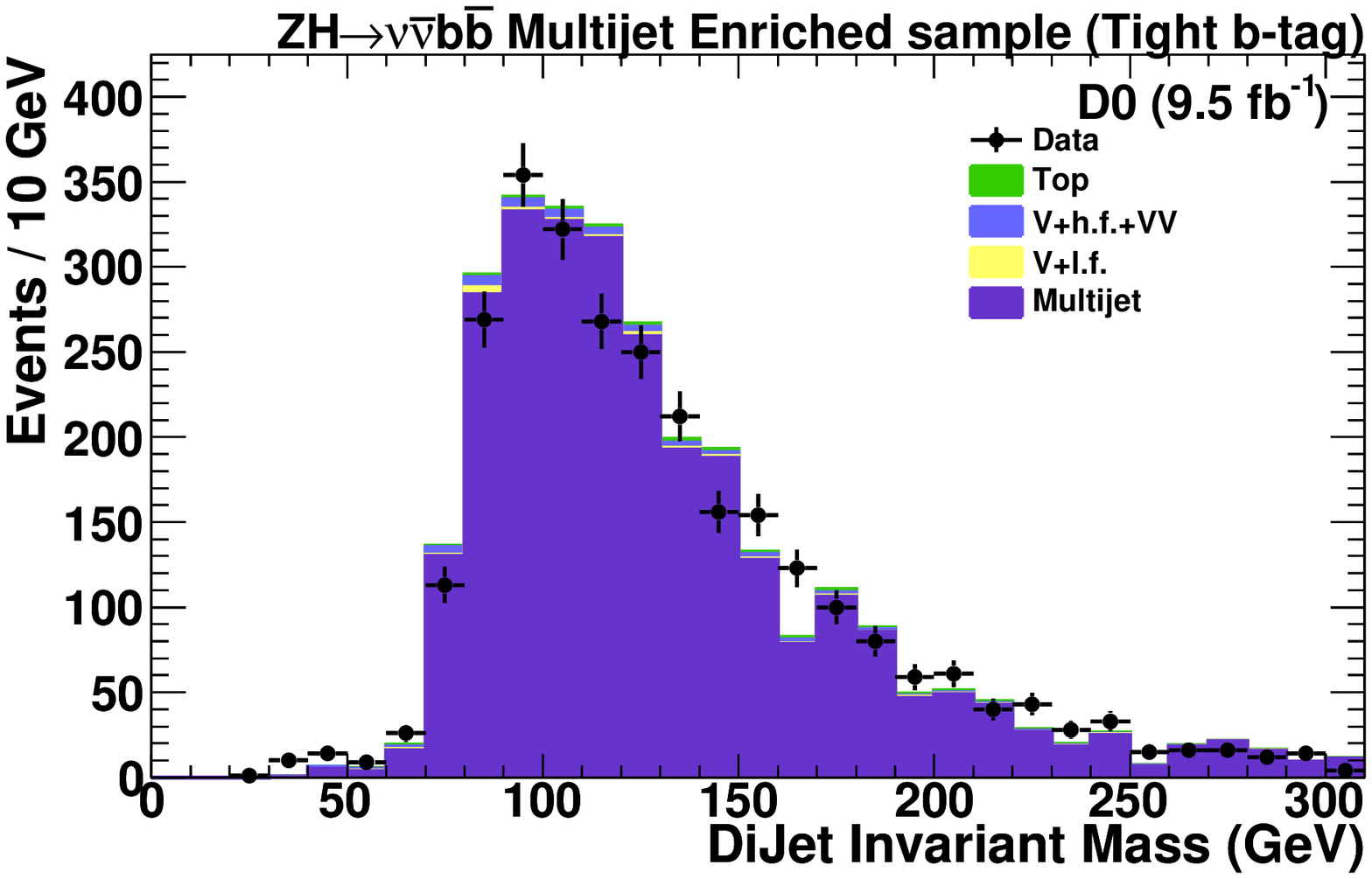}}
\caption{\label{JetDR_mj_pretag}
Representative variable distributions in the MJ-enriched sample:  
(a) dijet $\Delta R$ in the pre $b$-tag sample,
(b) $\mht/\hht$ (defined in Table~I) in the pre $b$-tag sample,
(c) dijet invariant mass in the medium $b$-tag sample,
(d) dijet invariant mass in the tight $b$-tag sample.
The data with ${\cal D} > \pi/2$ are shown as points and the 
background contributions as histograms: dibosons are labeled 
as ``VV,'' ``V+l.f.'' includes $(W/Z)$+$(u,d,s,g)$ jets, 
``V+h.f.'' includes $(W/Z)$+$(b,c)$ jets and ``Top'' includes 
pair and single top quark production. The ``multijet'' histogram 
is obtained from the data with ${\cal D} < \pi/2$
}
\end{figure*}

\bigskip
The distributions for the pre $b$-tag sample dijet $\Delta R$
and $\mht/\hht$ (defined in Table~\ref{tbl:DTvarsQCD}), and for 
the dijet invariant mass for medium $b$-tag and tight $b$-tag samples are shown 
in Fig.~\ref{JetDR_ew_pretag}
for the EW-control sample and 
in Fig.~\ref{JetDR_mj_pretag}
for the MJ-enriched sample.


\subsection{Analysis using decision trees}\label{sec:dt}

The full list of the seventeen input variables to the MJ DT is given in 
Table~\ref{tbl:DTvarsQCD}.

\bigskip
The MJ DT output is shown for the
analysis and EW control samples after the medium $b$ tagging requirement 
in Fig.~\ref{MJdecision}. 

\begin{figure*}[htp]
\centering
\subfigure[]{\includegraphics[width=8.5cm]{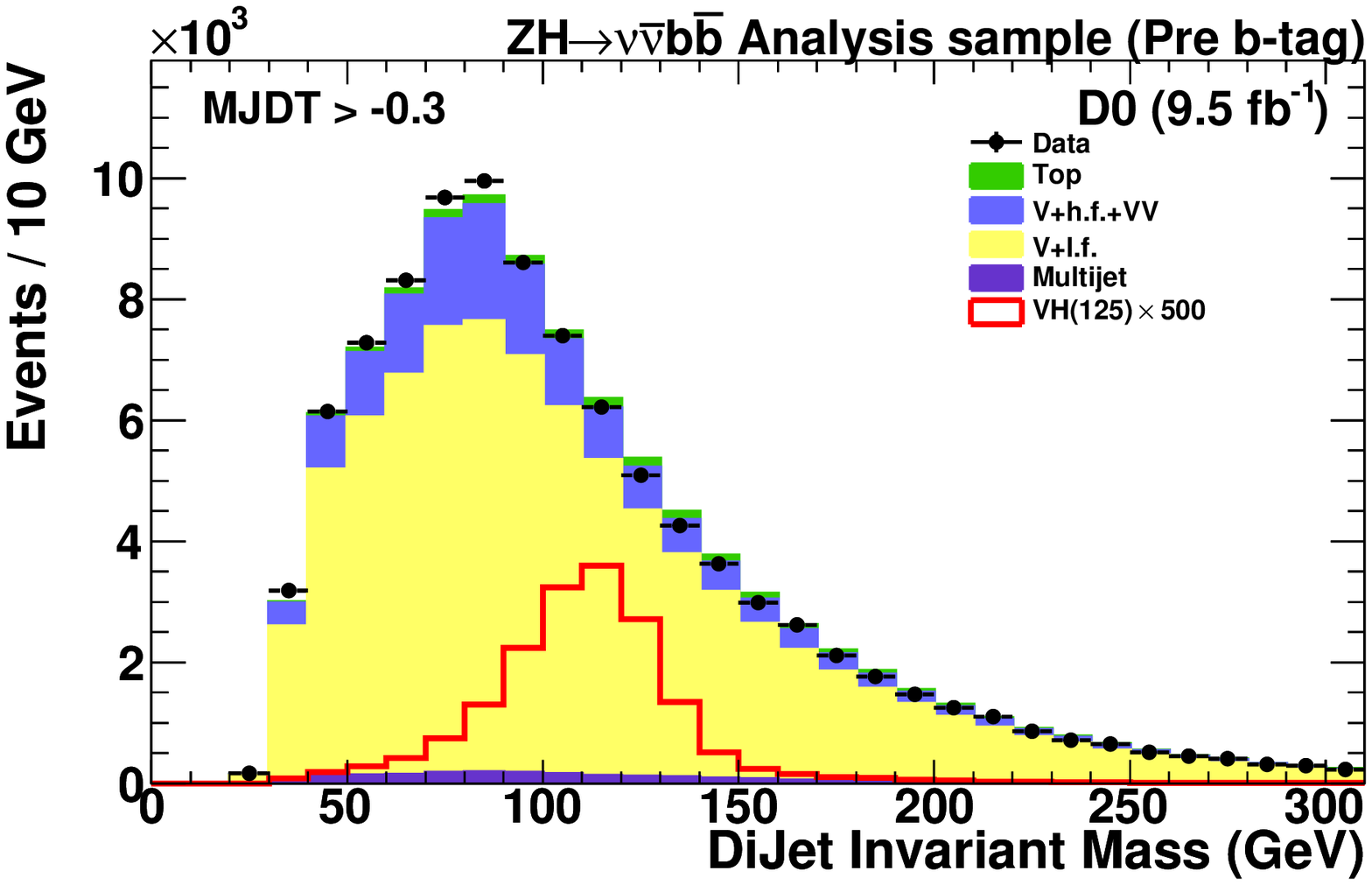}}
\subfigure[]{\includegraphics[width=8.5cm]{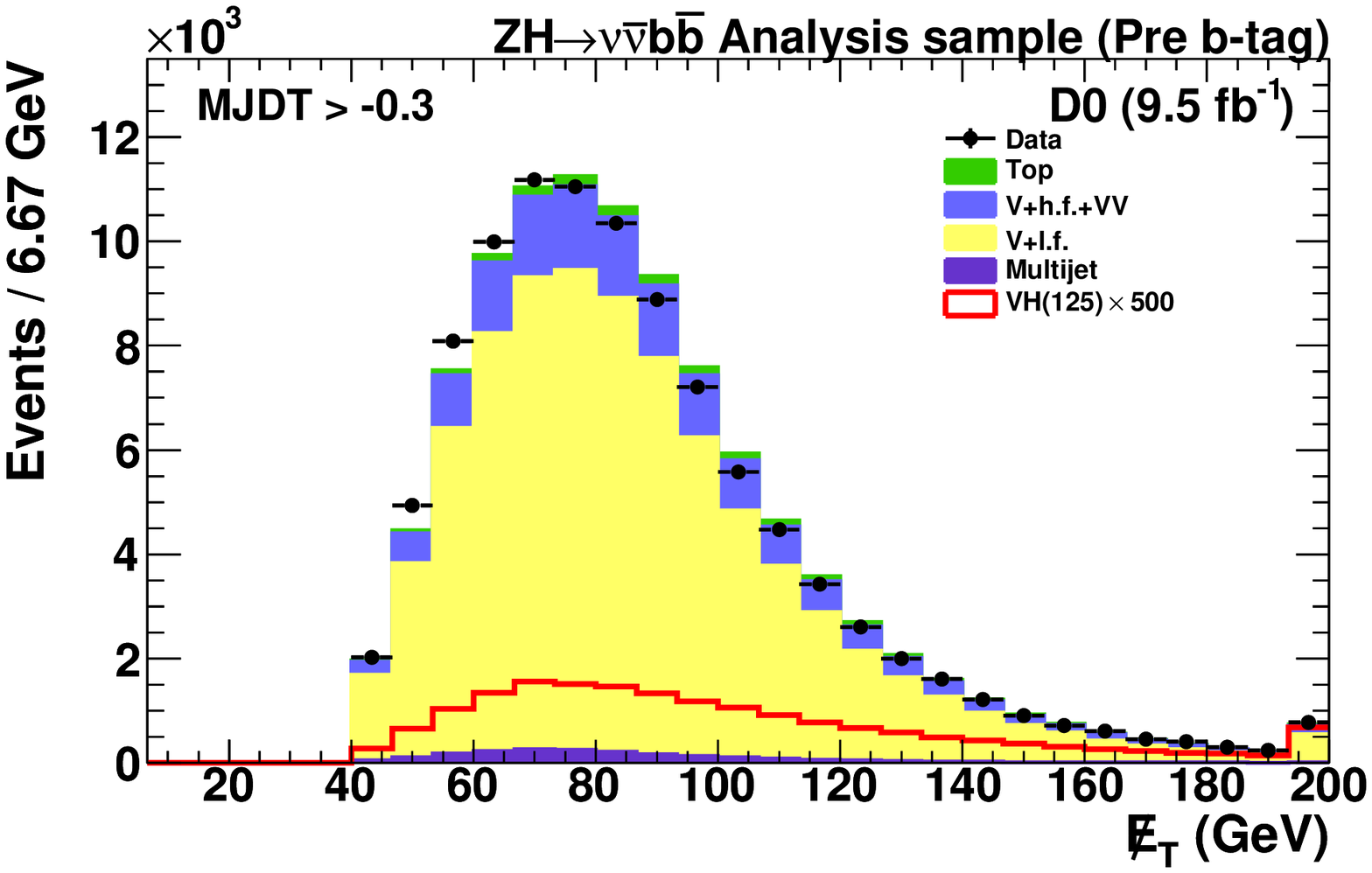}}
\subfigure[]{\includegraphics[width=8.5cm]{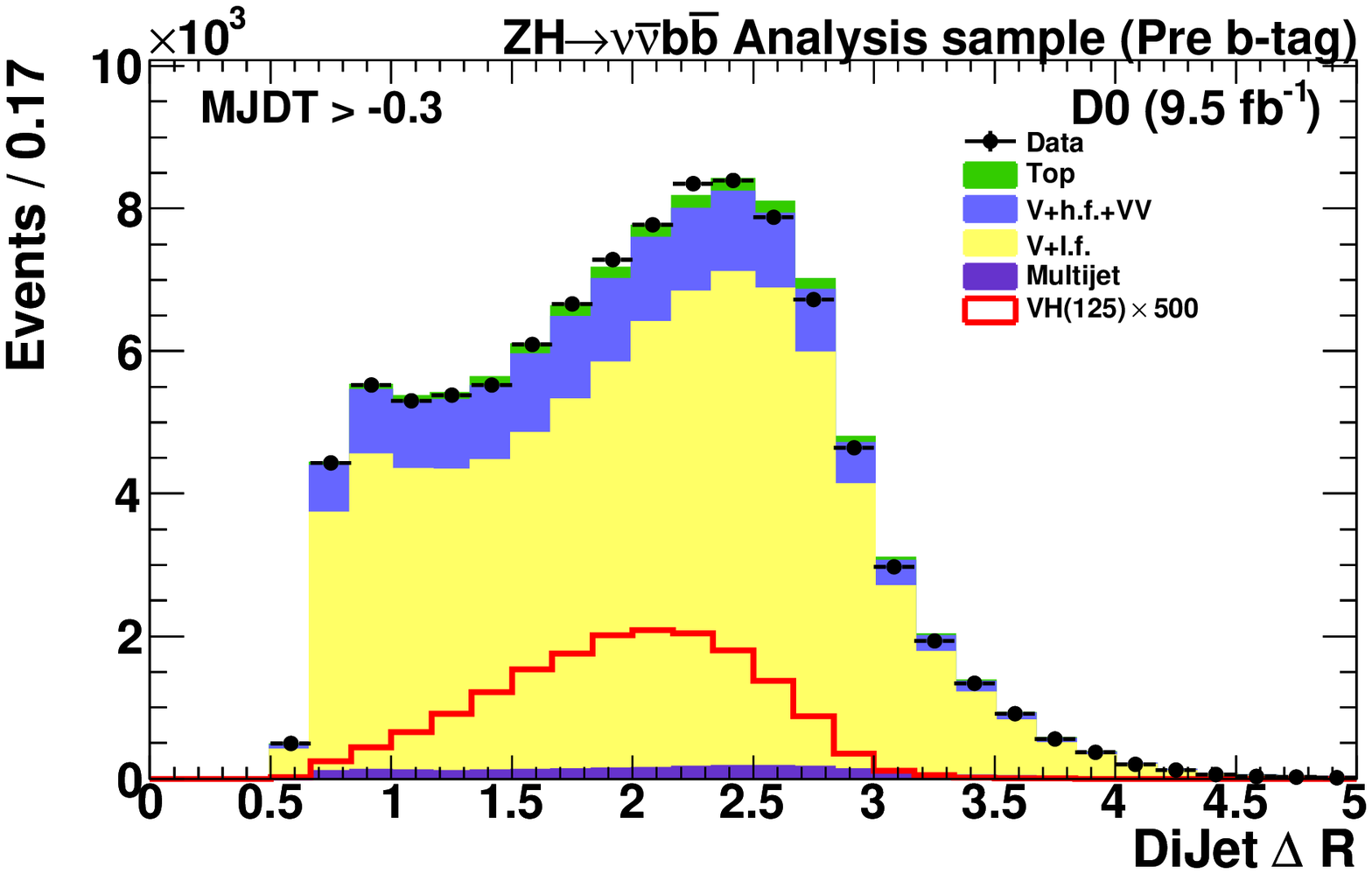}}
\subfigure[]{\includegraphics[width=8.5cm]{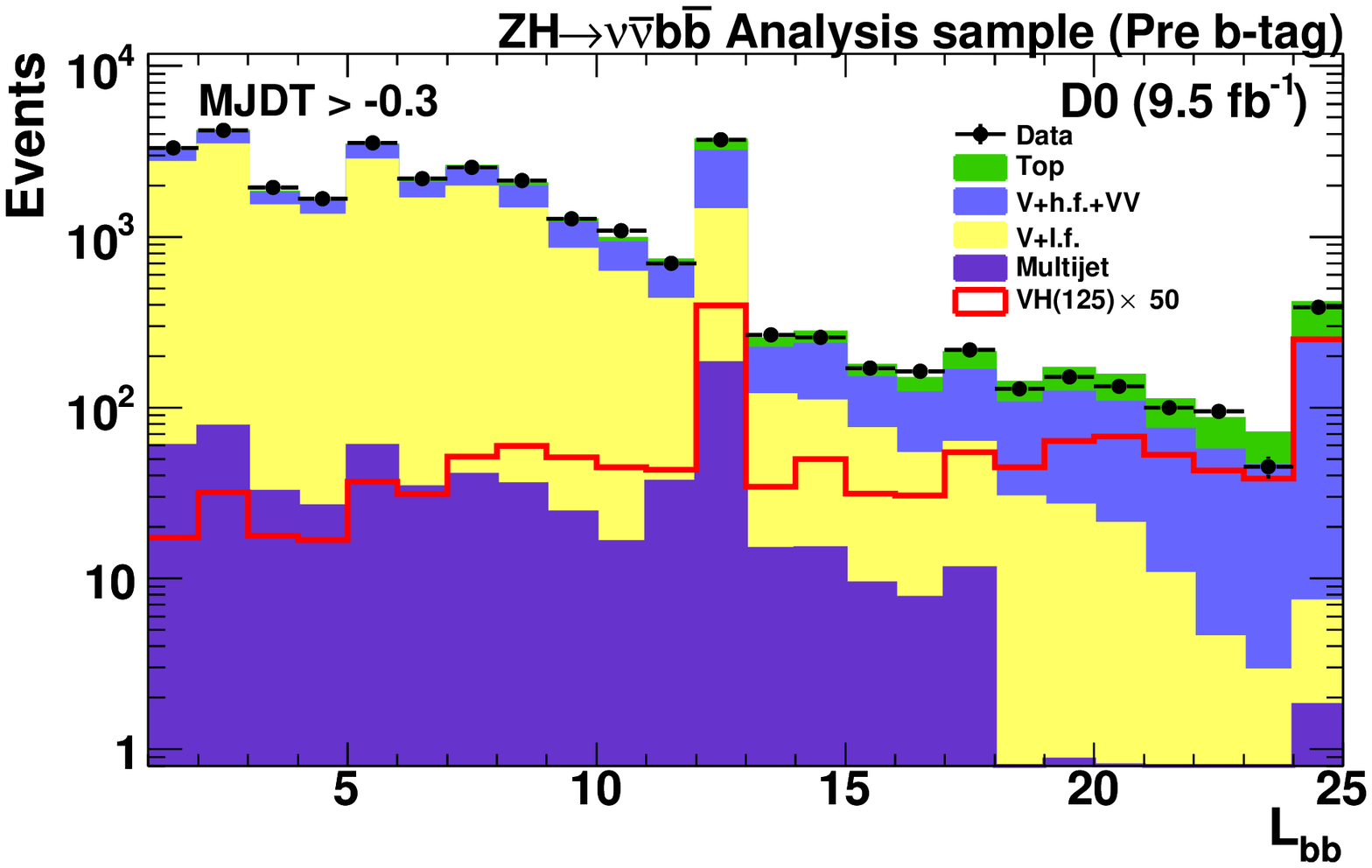}}
\caption{\label{JetDR_analysis_mjveto_pretag}
Representative variable distributions in the analysis sample after 
the multijet veto and before any $b$ tagging requirement:  
(a) dijet invariant mass,
(b) missing \et,
(c) dijet $\Delta R$,
(d) b-tagging discriminating variable ($L_{bb}$). The bin at zero is
surpressed in this plot due to the large number of entries, mostly
from pairs of light jets. The relatively high number of events observed 
at $L_{bb}=12$ comes mainly
from events with one untagged jet and one very tightly $b$-tagged jet;
the bin at $L_{bb}=24$ comes from events with two very tightly
$b$-tagged jets. The distributions for signal (VH), which are multiplied by a
factor of 500 for (a)--(c) and 50 for (d), include $ZH$ and $WH$ production 
for $m_H=125$~GeV.
The data are shown as points and the background contributions as histograms: 
dibosons are labeled as ``VV,'' ``V+l.f.'' includes $(W/Z)$+$(u,d,s,g)$ jets, 
``V+h.f.'' includes $(W/Z)$+$(b,c)$ jets and ``Top'' includes pair and single 
top quark production.}
\end{figure*}

\bigskip
The distributions for the dijet invariant mass, missing \et,
dijet $\Delta R$ and the b-tagging discriminating variable ($L_{bb}$)
are shown in Fig.~\ref{JetDR_analysis_mjveto_pretag} 
for the analysis sample after the multijet veto and before any
$b$-tagging requirement.

\bigskip
The full list of variables used in the SM DT is shown in 
Table~\ref{tbl:DTvarsQCD}. 


\subsection{Systematic uncertainties} \label{sec:systs}

Systematic uncertainties are summarized in Table~\ref{tab:systematics}. 
The numbers quoted are uncertainties on total yields. 
The background cross sections entry represents 
the global effect of cross section uncertainties on the sum of 
backgrounds. 
There is no luminosity uncertainty associated with the multijet normalization
since it comes from real data. The multijet is a non-negligible background 
component in medium $b$-tag sample.

In addition to the impact of these uncertainties on the integrated signal 
and background yields, modifications of the shapes of the final
discriminants are also considered, when relevant. These originate mainly
from jet corrections (energy scale, resolution and $b$-tagging) and also
have small contributions from Monte Carlo reweightings and from parton
distribution function variations.


\subsection{Limit setting procedure} 

Figure~\ref{bkgdsubtract} shows for $m_H=125$~GeV the
SM DT distributions after profiling. In this case, the background
prediction and its uncertainties have been determined from the fit to
data under the background-only hypothesis.


\subsection{Higgs Search Results}

No added material.


\subsection{Diboson Search Results}

\begin{figure*}[htp]
\centering
\subfigure[]{\includegraphics[width=8.5cm]{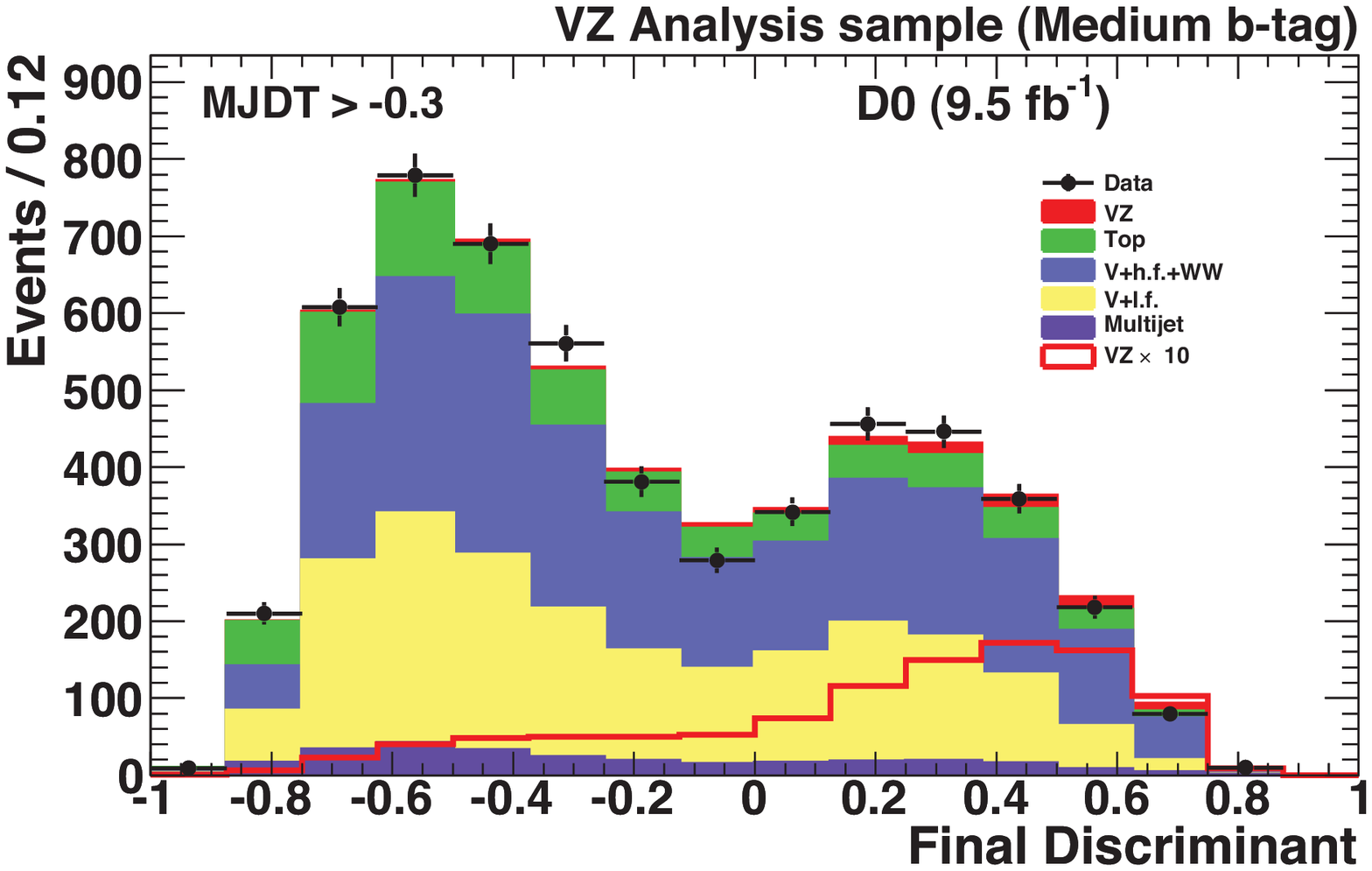}}
\subfigure[]{\includegraphics[width=8.5cm]{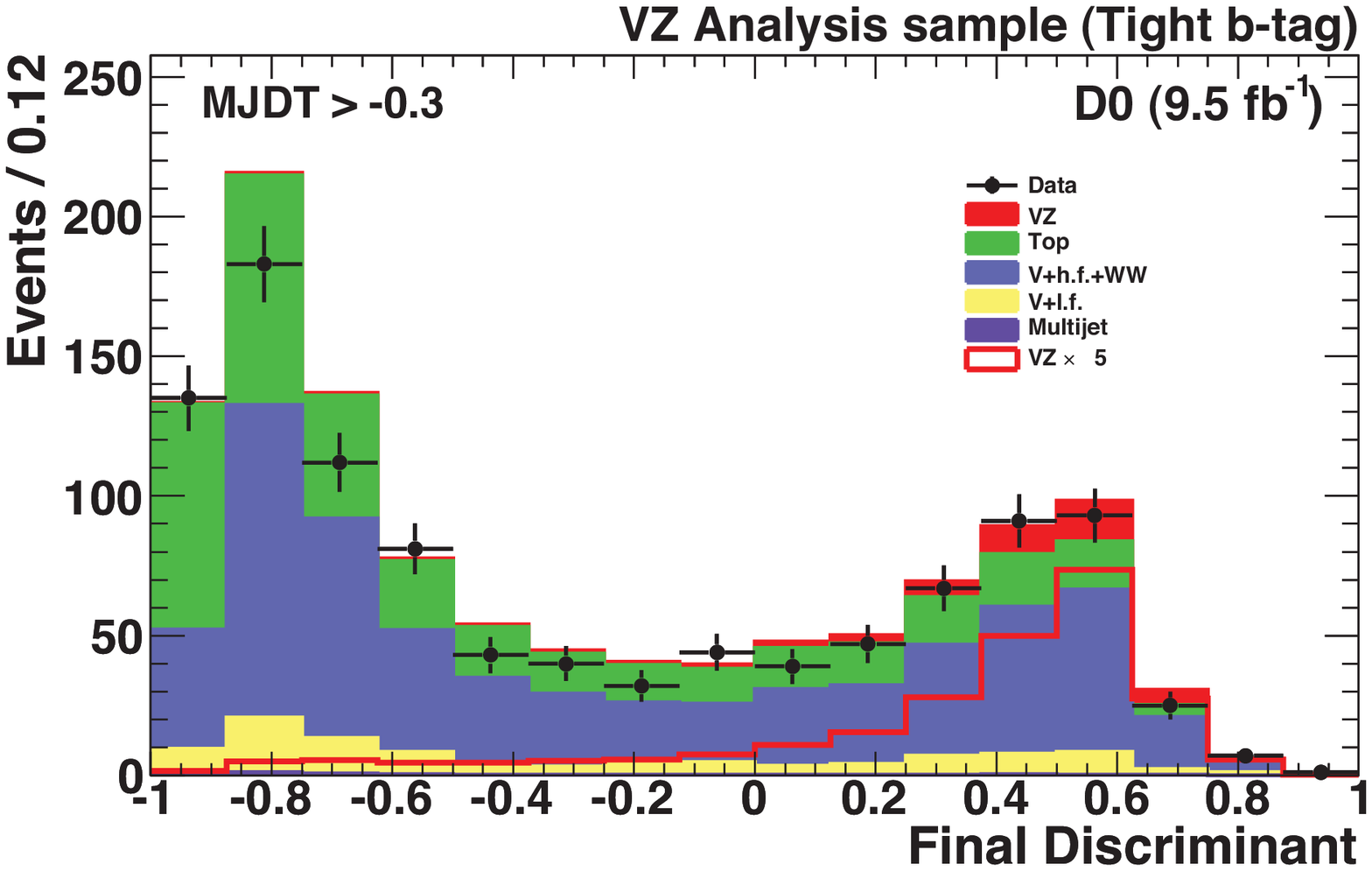}}
\caption{\label{DT_Diboson}
The SM DT output for the $WZ$ and $ZZ$ diboson search following 
the multijet veto
for (a) medium and (b) tight tag prior to the fit to data. 
The data are shown as points 
and the background contributions as histograms; ``V+l.f.'' 
includes $(W/Z)$+$(u,d,s,g)$ jets, 
``V+h.f.'' includes $(W/Z)$+$(b,c)$ jets and ``Top'' includes 
pair and single top quark production.
The $WZ$ and $ZZ$ signal is denoted as VZ. 
The distributions for signal are scaled to the SM cross section 
(filled red histogram) and shown separately multiplied by a
factor of 10 for medium $b$-tag and 5 for tight $b$-tag (solid red line) 
respectively.}
\end{figure*}

The medium and tight $b$-tag SM DTs are shown in Fig.~\ref{DT_Diboson}. 

\bigskip
Figure~\ref{bkgdsubtract_diboson_sbfit} shows the SM DT distributions.
The background prediction and its uncertainties have been 
determined from a fit to the data under the signal+background hypothesis.  

\begin{figure*}[htp]
\centering
\subfigure[]{\includegraphics[width=8.5cm]{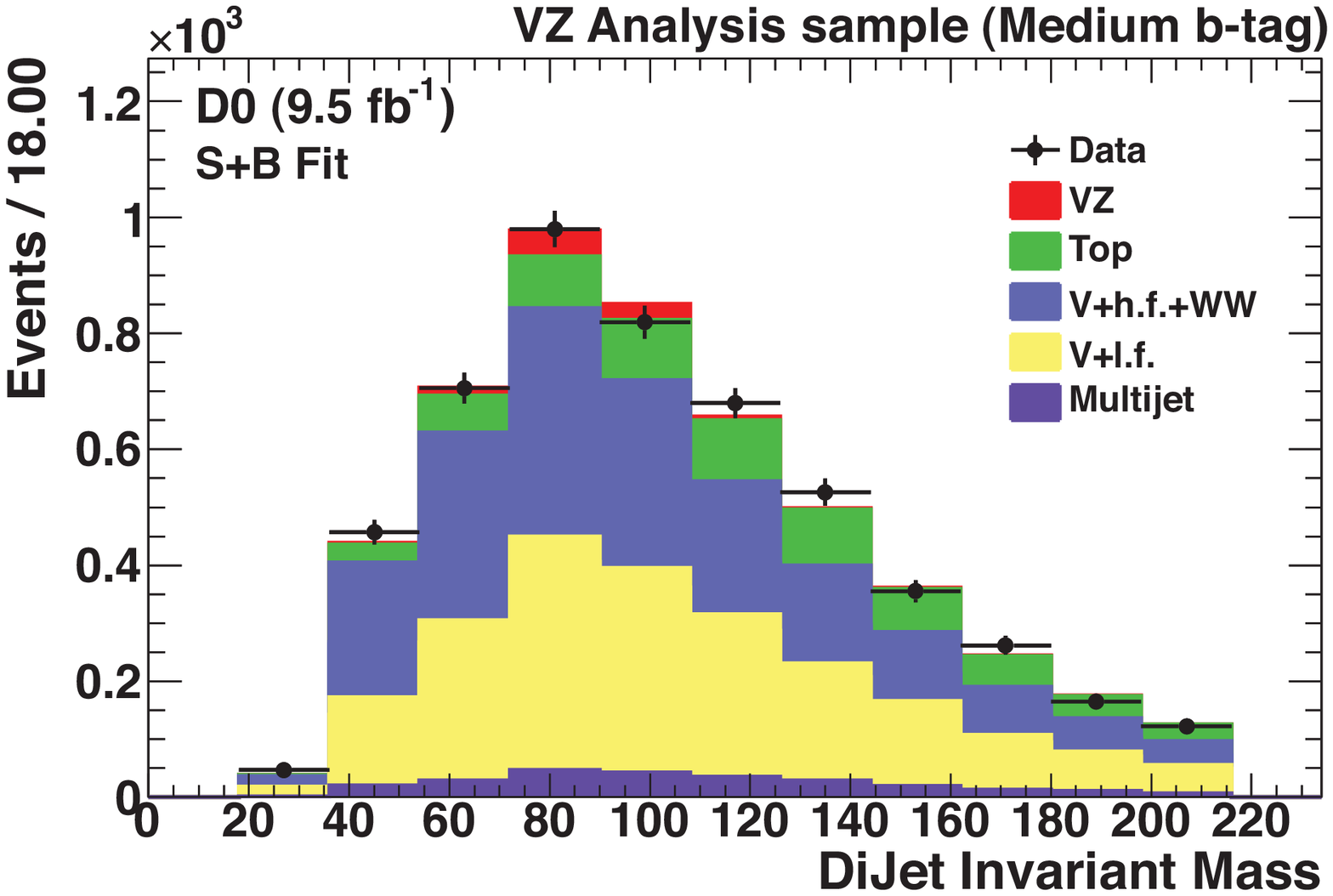}}
\subfigure[]{\includegraphics[width=8.5cm]{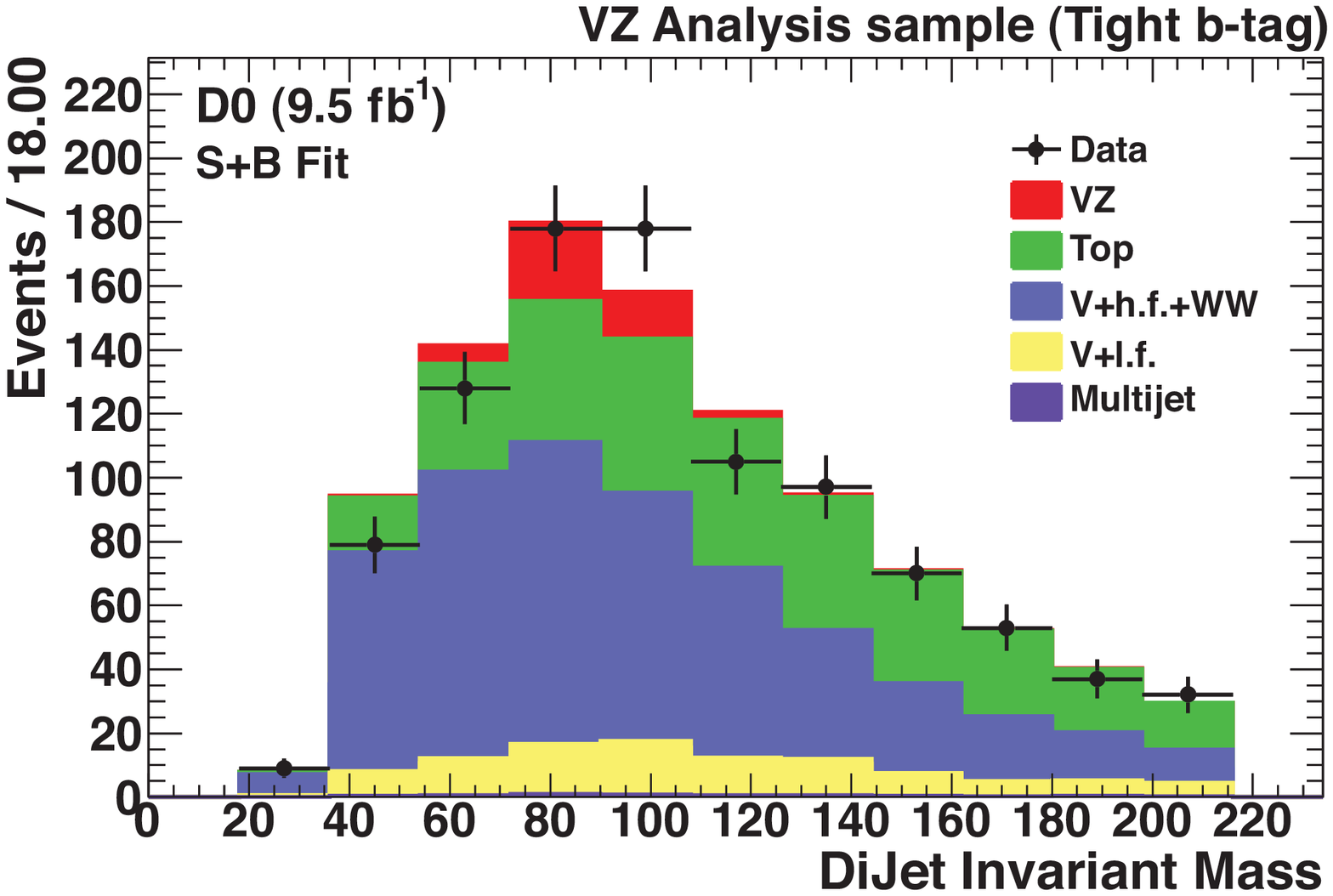}}
\subfigure[]{\includegraphics[width=8.5cm]{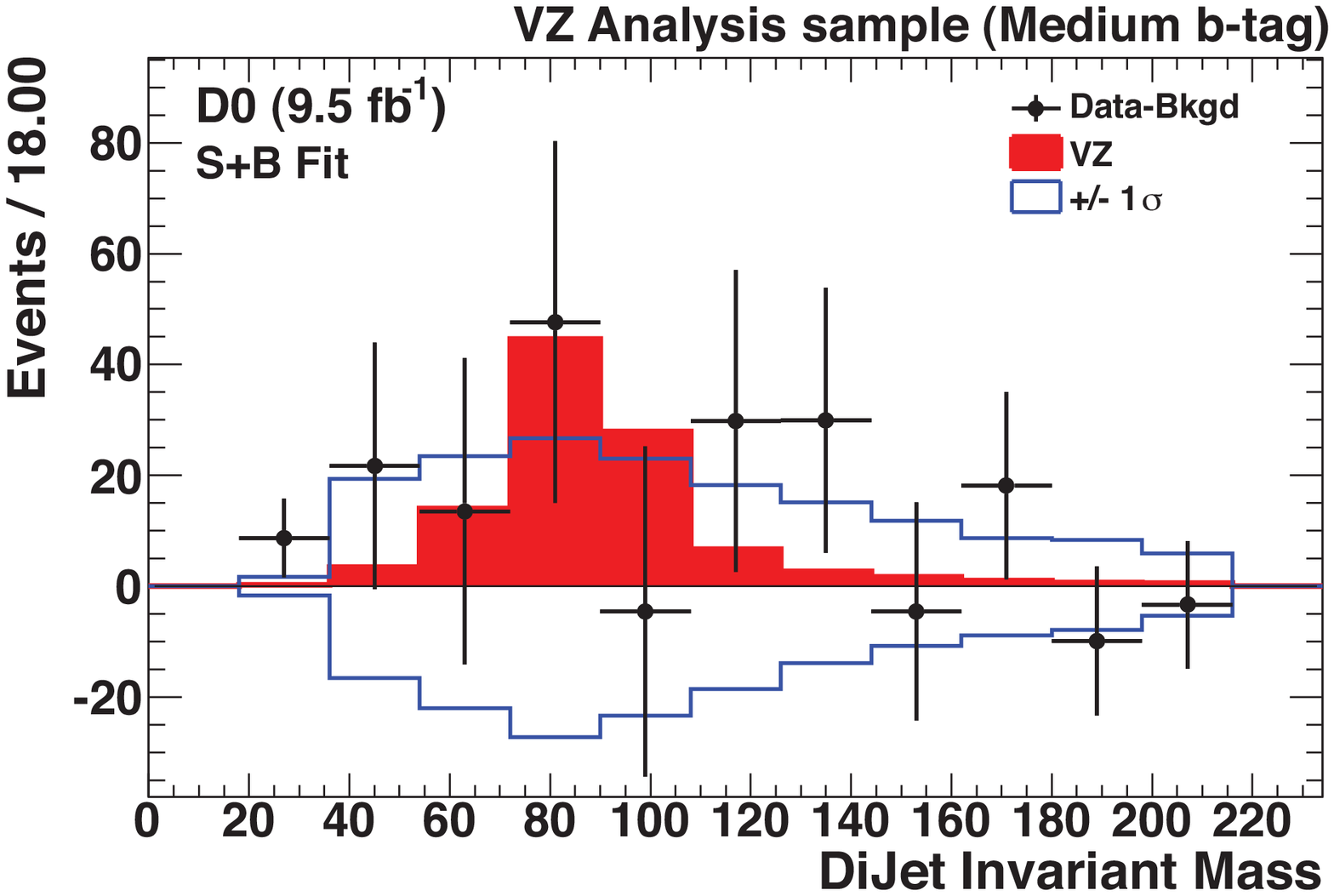}}
\subfigure[]{\includegraphics[width=8.5cm]{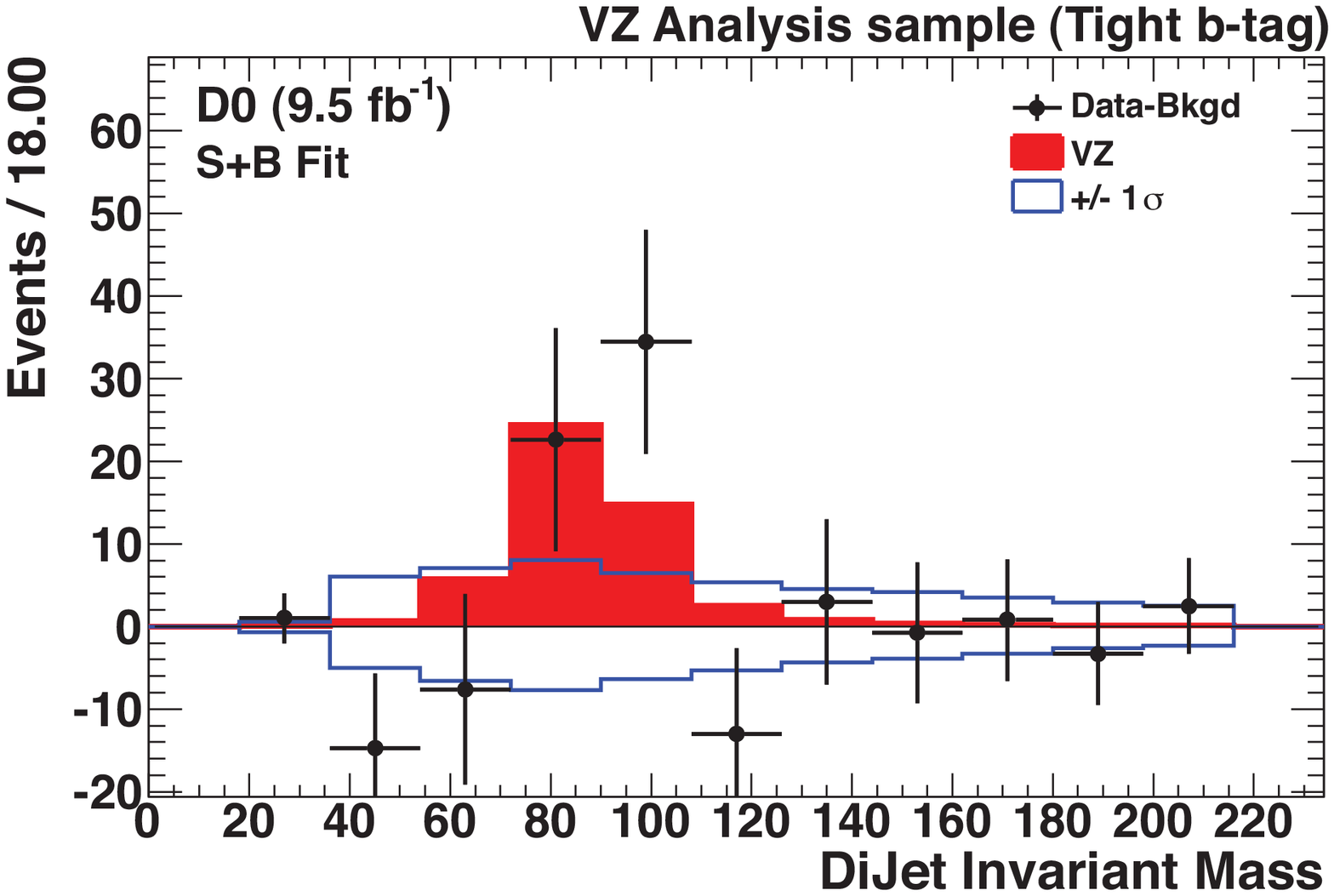}}
\caption{\label{bkgdsubtract_diboson_dim_sbfit}
The dijet invariant mass for the $WZ$ and $ZZ$ diboson search, 
following the multijet veto, 
and after the fit to the data under the signal+background 
hypothesis in the (a) medium and (b) tight tag channels. 
The data are shown as points and the background contributions as histograms;
 ``V+l.f.'' includes $(W/Z)$+$(u,d,s,g)$ jets,
 ``V+h.f.'' includes $(W/Z)$+$(b,c)$ jets 
and ``Top'' includes pair and single top quark production. 
The $WZ$ and $ZZ$ signal expectation (red histogram, and denoted VZ) and 
the data after subtracting the fitted background (points) are shown in the 
(c) medium and (d) tight tag channels.
Also shown is the $\pm 1$ standard deviation band on the total background 
after fitting. The signal is scaled to the SM cross section.
}
\end{figure*}

\bigskip
Figure~\ref{bkgdsubtract_diboson_dim_sbfit} shows 
the dijet invariant mass distributions,
along with the background-subtracted data. The background 
prediction and its uncertainties have 
been determined from a fit to the data under the signal+background hypothesis.


\subsection{Summary}

No added material.


\begin{table*}
\caption{\label{tbl:DTvarsQCD} Variables used as input to the decision
trees, where the angles $\theta$ and $\phi$ are the polar and
azimuthal angles defined with respect to the proton beam
direction. $\mathrm{jet_1}$ refers to the leading taggable jet,
$\mathrm{jet_2}$ refers to the next-to-leading taggable jet,
$j_{\mathrm{all}}$ refers to any jet in the event with
$\pt>15$~GeV, pseudorapidity $\vert\eta\vert<3.2$ and without the 
taggability requirement. 
The thrust axis is the direction obtained from the difference of the
  transverse momenta of the leading and next-to-leading jets. The
recoil is defined in the plane transverse to the beam using i) either
the amount of missing transverse energy that remains after removal of
the two leading jets, ii) or the sum of all good jet transerse momenta
in the half plane opposite to the one containing the dijet system (with
respect to the thrust axis). Among these two possible recoil
definitions, the one that has the larger component
along the direction orthogonal to the thrust is chosen.}

\begin{center}
\begin{tabular}{l}
\hline
\hline
\centerline{Variables used in the MJ DT} \\
\hline
$\Delta\phi\jj$\\
$\eta$ of $\mathrm{jet_1}$\\
\met \\
\met\ significance\\
$\min\Delta\phi(\met,j_{\mathrm{all}})$\\
$\max\Delta\phi(\met,j_{\mathrm{all}})+\min\Delta\phi(\met,j_{\mathrm{all}})$\\
$\max\Delta\phi(\met,j_{\mathrm{all}})-\min\Delta\phi(\met,j_{\mathrm{all}})$\\
\mht (vectorial sum of $j_{\mathrm{all}}$ \pt)\\
\mht / \hht\ (with \hht\ the scalar sum of $j_{\mathrm{all}}$ \pt)\\
Asymmetry between \met\ and \mht: $(\met-\mht)/(\met+\mht)$\\
\met\ component along the thrust axis\\
\met\ component perpendicular to the thrust axis\\
Sum of the signed components of the dijet and recoil momenta along the
thrust axis \\
Sum of the signed components of the dijet and recoil momenta
perpendicular to the thrust axis\\
Centrality (ratio of the scalar sum of $\mathrm{jet_1}$ and
$\mathrm{jet_2}$ \pt\ to the sum of their energies)\\
$\theta$ angle of the dijet system\\
Polar angle of $\mathrm{jet_1}$ boosted to the dijet rest frame with 
respect to the dijet direction in the laboratory \\
\hline
\hline
\centerline{Variables used in the SM DT}\\
\hline
Dijet mass\\
Dijet transverse mass\\
$\mathrm{jet_1}$ \pt\\
$\mathrm{jet_2}$ \pt\\
Scalar sum of $\mathrm{jet_1}$ and $\mathrm{jet_2}$ \pt\\
$\eta$ of $\mathrm{jet_1}$\\
$\eta$ of  $\mathrm{jet_2}$\\
$\Delta\eta\jj$\\
$\Delta\phi\jj$\\
$\Delta R\jj$\\
$p_T$ weighted $\Delta R(\mathrm{jet_1},j_{\mathrm{all}})$\\
$p_T$ weighted $\Delta R(\mathrm{jet_2},j_{\mathrm{all}})$\\
\hht (scalar sum of $j_{\mathrm{all}}$ \pt)\\
\mht (vectorial sum of $j_{\mathrm{all}}$ \pt)\\
\mht / \hht \\
$\Delta\phi(\met ,\mathrm{dijet})$\\
$\theta$ angle of $\mathrm{jet_1}$ boosted to the dijet rest frame\\
Polar angle of $\mathrm{jet_1}$ boosted to the dijet rest frame 
with respect to the dijet direction in the laboratory \\
$\min\Delta\phi(\met,j_{\mathrm{all}})$\\
$\max\Delta\phi(\met,j_{\mathrm{all}})+\min\Delta\phi(\met,j_{\mathrm{all}})$\\
Dijet \pt \\
$\Delta\phi(\met ,\mathrm{jet_1})$\\
\hline
\hline
\end{tabular}
\end{center}
\end{table*}


\begin{table*}
\caption{\label{tab:systematics}
Systematic uncertainties, in percent, of the overall 
signal and background yields. 
``Jet EC'' and ``Jet ER'' stand for jet 
energy calibration and resolution respectively. 
``Jet R\&T'' stands for jet reconstruction and taggability.
``Signal'' includes $ZH$ 
and $WH$ production and is shown for $m_H=125$~GeV.}
\begin{center}
\begin{tabular}{l|c|c}
\hline
\hline
Systematic Uncertainty & Signal (\%) & Background (\%) \\
\hline
\multicolumn{3}{c}{Medium $b$-tag} \\
\hline
Jet EC - Jet ER & $\pm$ 0.9 & $\pm$ 1.9 \\
Jet R\&T & $\pm$ 2.9 & $\pm$ 2.9 \\
b Tagging & $\pm$ 0.6 & $\pm$ 3.7 \\
Trigger & $\pm$ 2.0 & $\pm$ 1.9 \\
Lepton Identification & $\pm$ 0.8 & $\pm$ 0.9 \\
Heavy Flavor Fractions & $-$ & $\pm$ 8.5 \\
Cross Sections & $\pm$ 7.0 & $\pm$ 9.8 \\
Luminosity & $\pm$ 6.1 & $\pm$ 5.8 \\
Multijet Normalilzation & $-$ & $\pm$ 1.2 \\
Total & $\pm$ 10.0 & $\pm$ 14.2 \\ 
\hline
\multicolumn{3}{c}{Tight $b$-tag} \\
\hline
Jet EC - Jet ER & $\pm$ 1.0 & $\pm$ 1.8 \\
Jet R\&T & $\pm$ 2.7 & $\pm$ 3.1 \\
b Tagging & $\pm$ 8.6 & $\pm$ 7.4 \\
Trigger & $\pm$ 2.0 & $\pm$ 2.0 \\
Lepton Identification & $\pm$ 0.9 & $\pm$ 1.1 \\
Heavy Flavor Fractions & $-$ & $\pm$ 11.1 \\
Cross Sections & $\pm$ 7.0 & $\pm$ 10.0 \\
Luminosity & $\pm$ 6.1 & $\pm$ 6.1 \\
Multijet Normalilzation & $-$ & $\pm$ 0.1 \\
Total & $\pm$ 13.2 & $\pm$ 16.9 \\ 
\hline
\hline
\end{tabular}
\end{center}
\end{table*}


\begin{figure*}[htp]
\centering
\subfigure[]{\includegraphics[width=8.5cm]{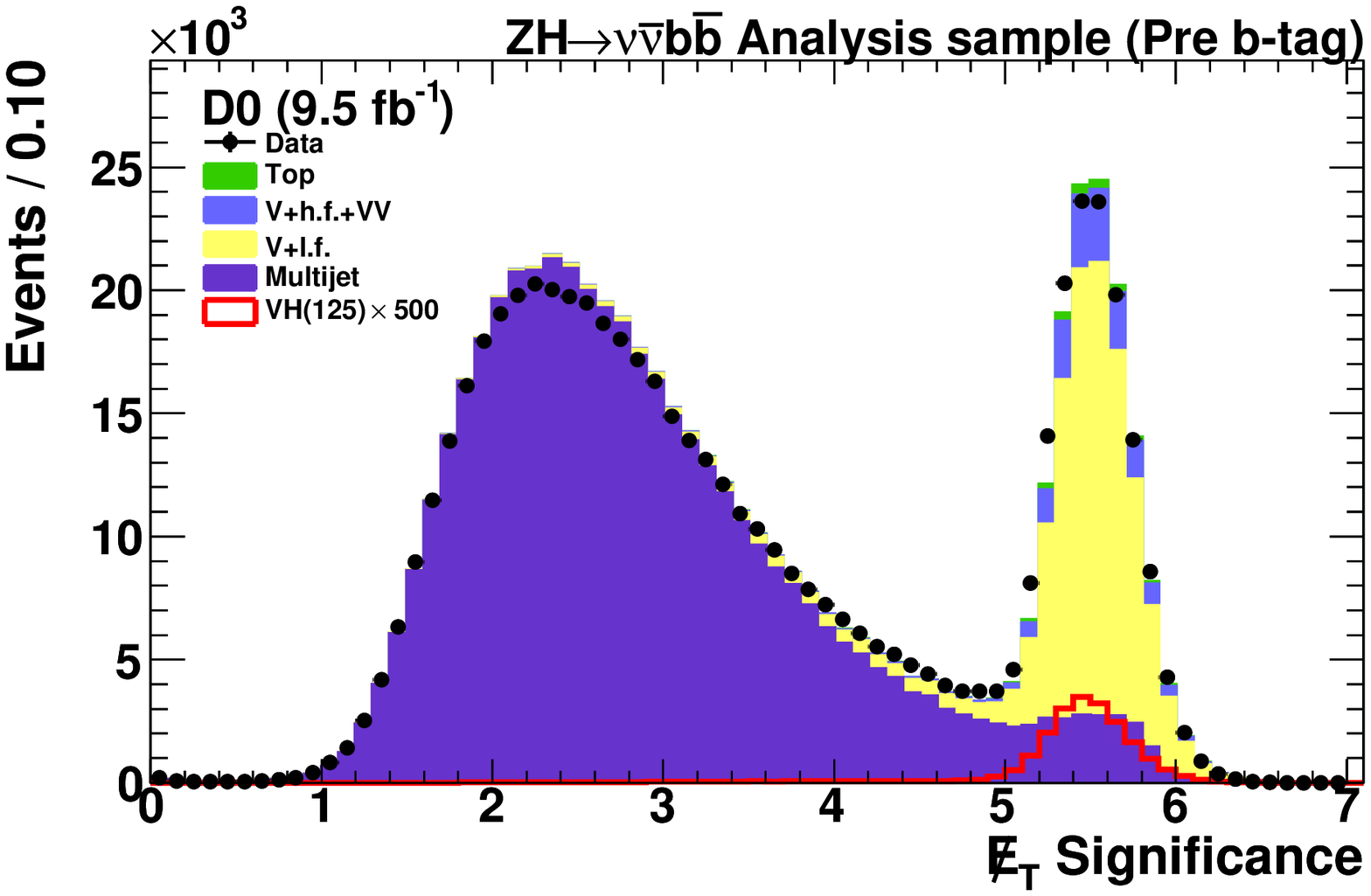}}
\subfigure[]{\includegraphics[width=8.5cm]{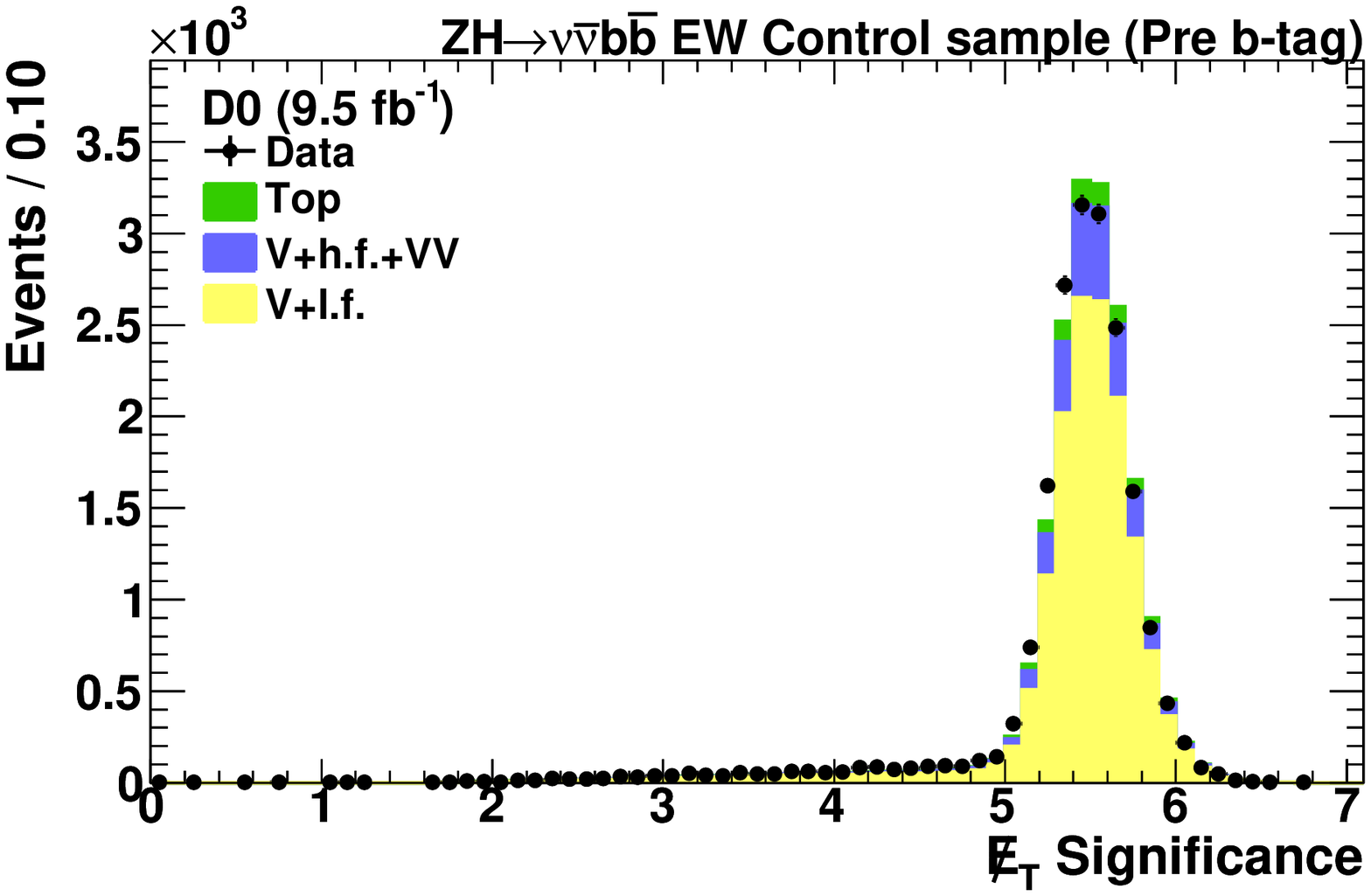}}
\caption{\label{HM_analysis_pretag}  
Missing \et\ significance in (a) the analysis and (b) the EW-control
samples without the requirement that the significance be larger than 5.
The data are shown as points and the background contributions as histograms: 
dibosons are labeled as ``VV,'' ``V+l.f.'' includes $(W/Z)$+$(u,d,s,g)$ jets, 
``V+h.f.'' includes $(W/Z)$+$(b,c)$ jets and ``Top'' includes pair and single 
top quark production. In (a),
the distribution for signal (VH) is multiplied by a factor of 500 and 
includes $ZH$ and $WH$ production for $m_H=125$~GeV.
}
\end{figure*}

\vfill

\begin{figure*}[tp]
\centering
\subfigure[]{\includegraphics[width=8.5cm]{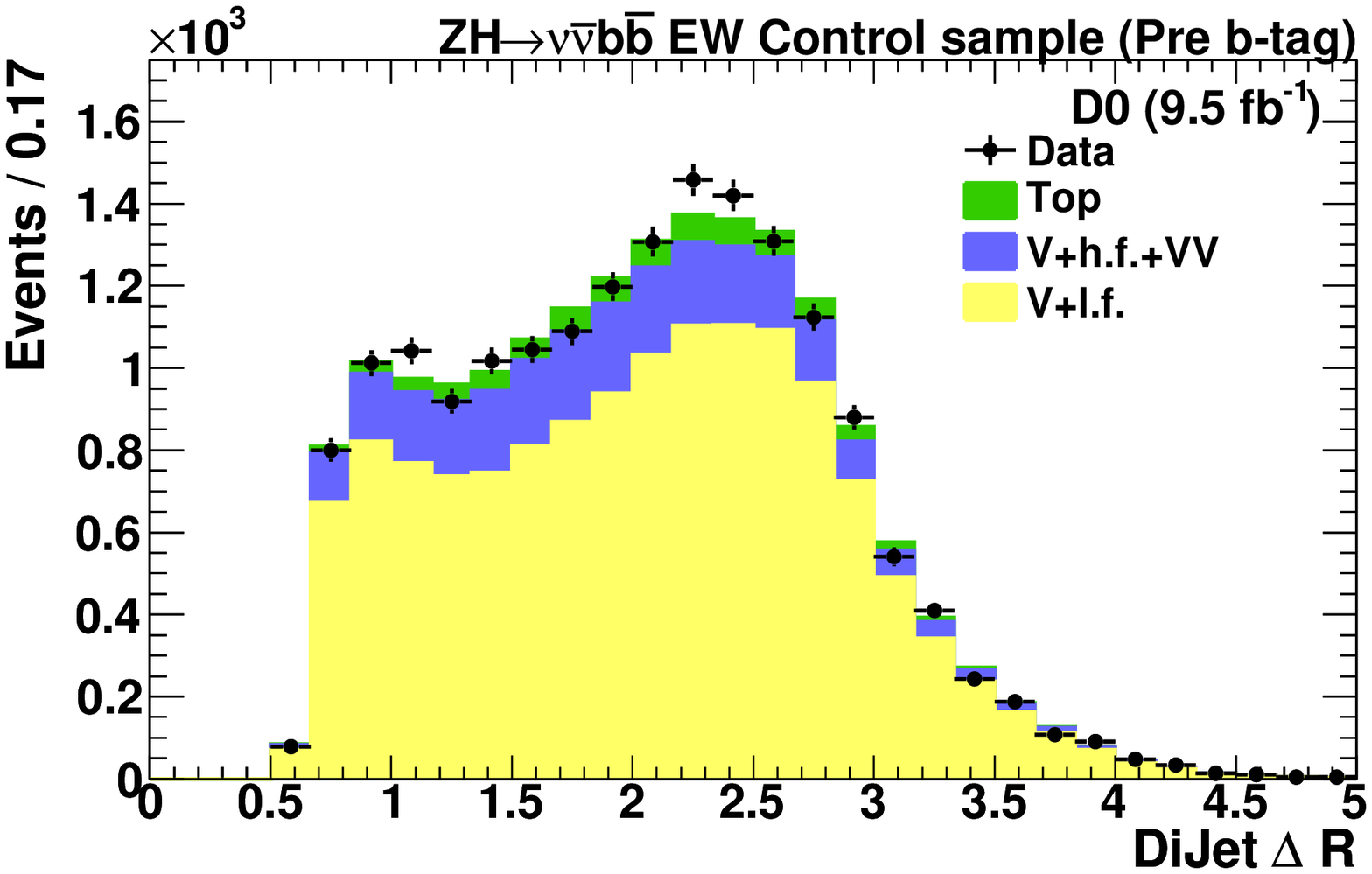}}
\subfigure[]{\includegraphics[width=8.5cm]{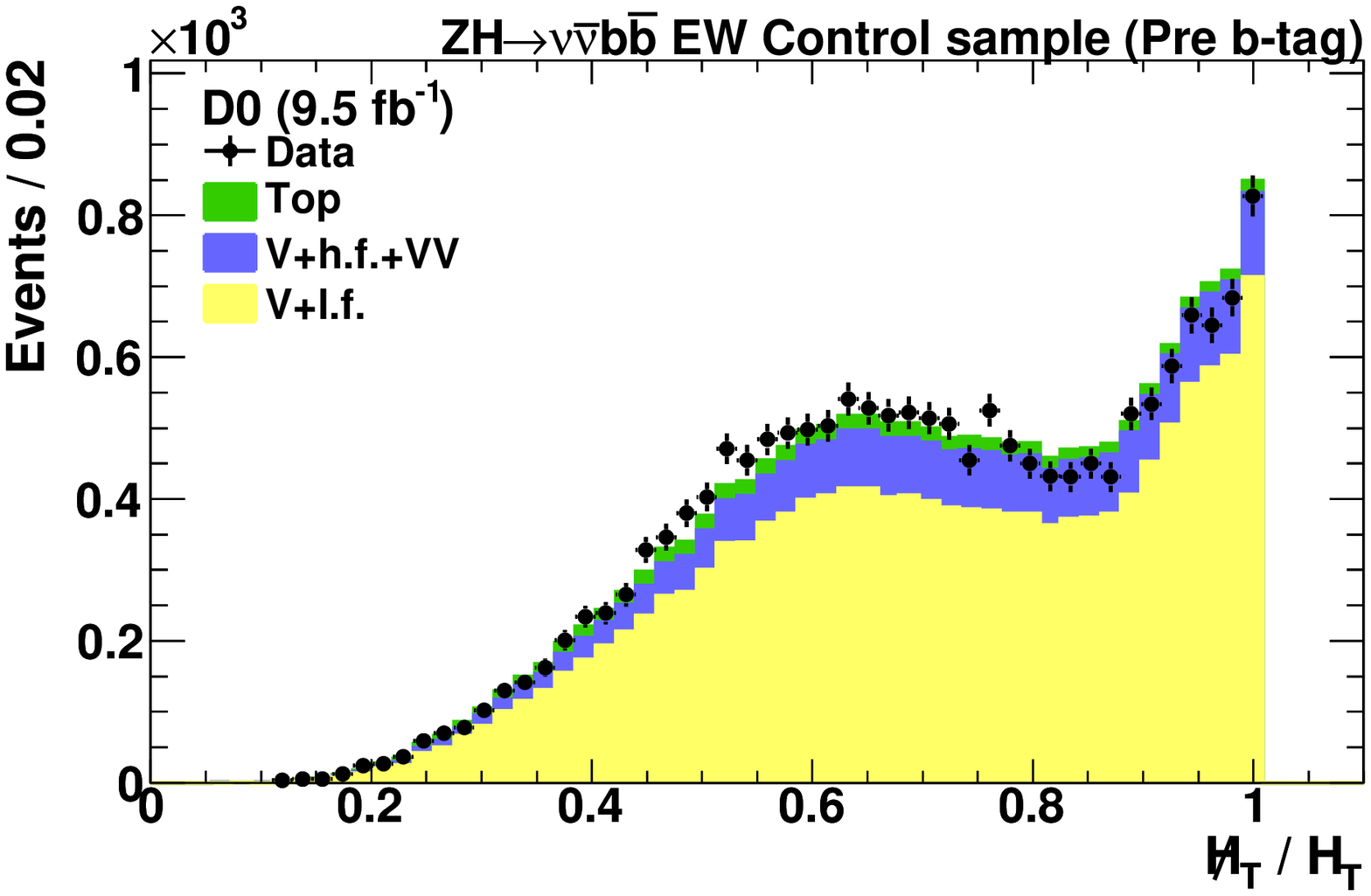}}
\subfigure[]{\includegraphics[width=8.5cm]{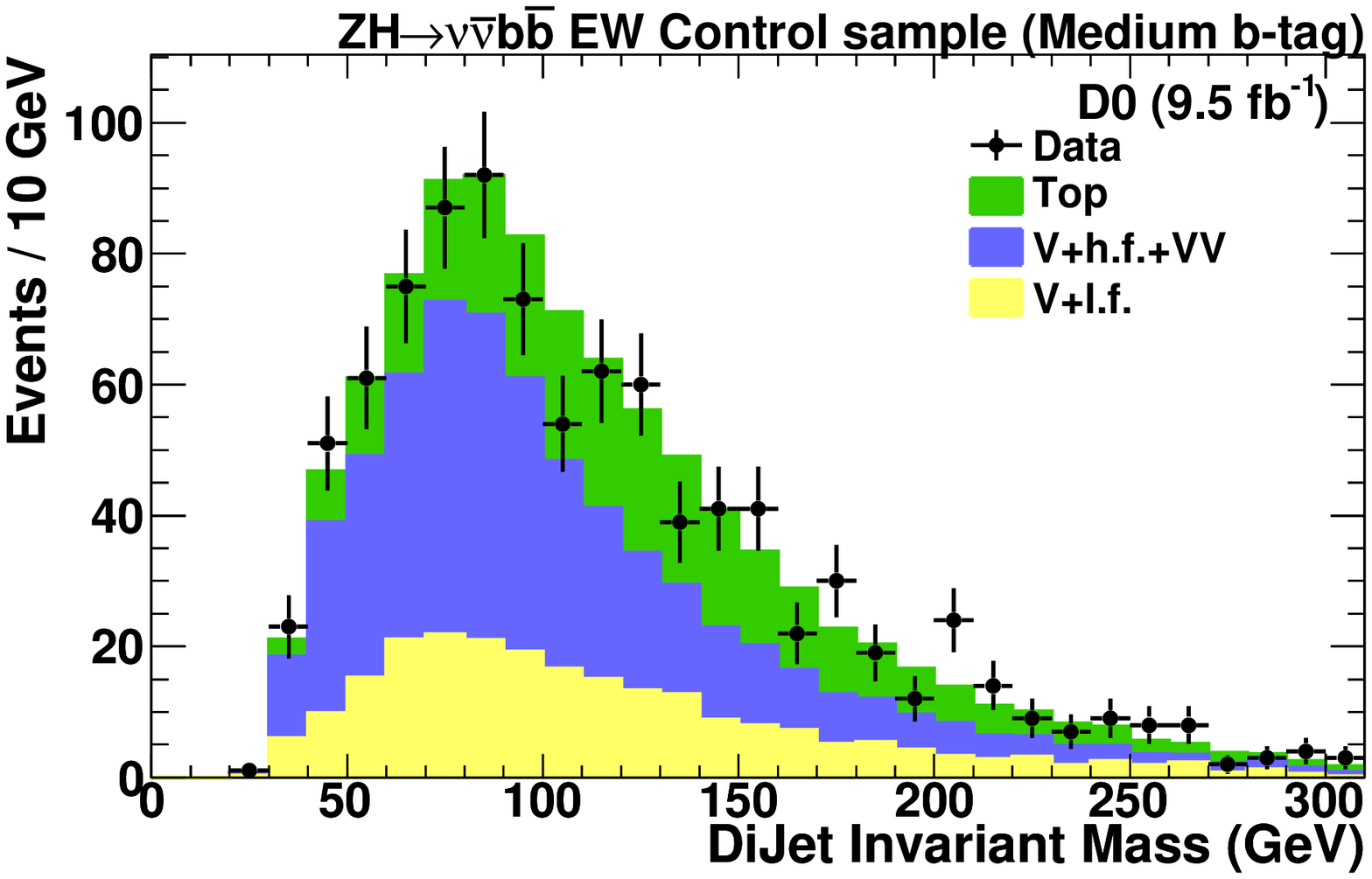}}
\subfigure[]{\includegraphics[width=8.5cm]{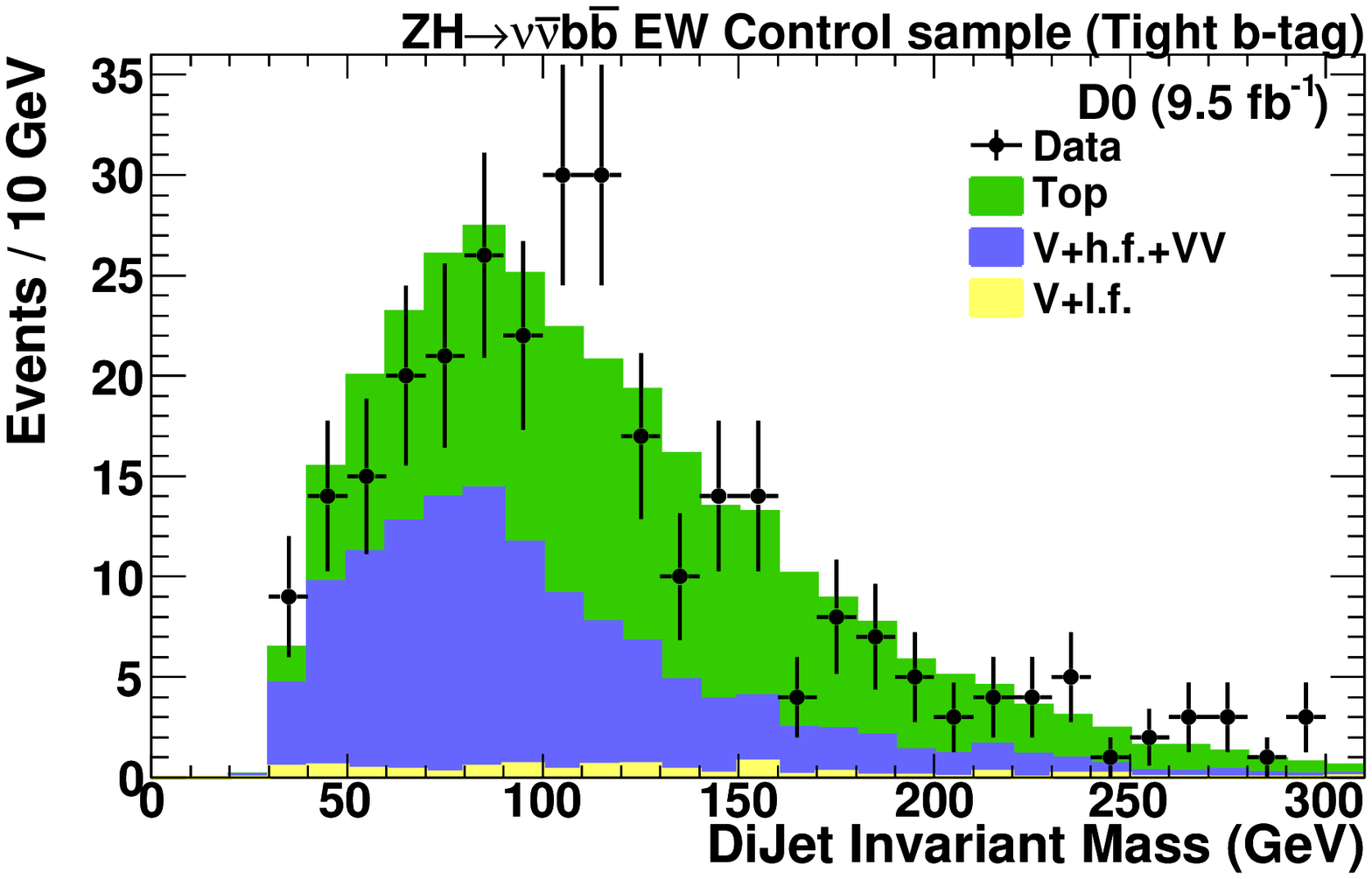}}
\caption{\label{JetDR_ew_pretag}
Representative variable distributions in the EW-control sample:  
(a) dijet $\Delta R$ in the pre $b$-tag sample,
(b) $\mht/\hht$ (defined in Table~I) in the pre $b$-tag sample,
(c) dijet invariant mass in the medium $b$-tag sample,
(d) dijet invariant mass in the tight $b$-tag sample.
The data are shown as points and the background contributions as histograms: 
dibosons are labeled as ``VV,'' ``V+l.f.'' includes $(W/Z)$+$(u,d,s,g)$ jets, 
``V+h.f.'' includes $(W/Z)$+$(b,c)$ jets and ``Top'' includes pair and single 
top quark production.}
\end{figure*}

\begin{figure*}[htp]
\centering
\begin{tabular}{cc}
\subfigure[]{\includegraphics[width=8.5cm]{Signal_DT_QCD_1tag.eps}}
\subfigure[]{\includegraphics[width=8.5cm]{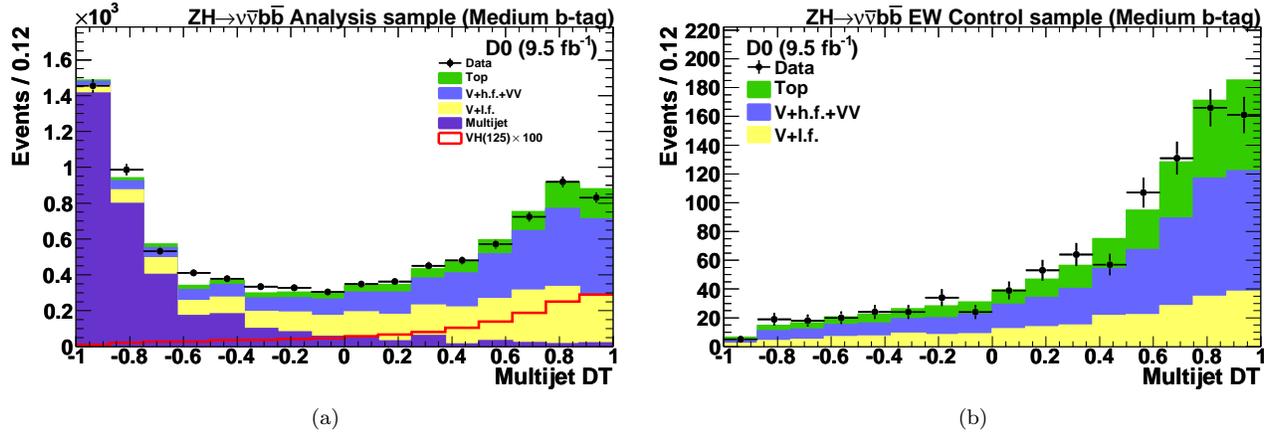}}
\end{tabular}
\caption{\label{MJdecision}
MJ DT output after the medium $b$-tagging requirement in the (a) analysis
sample and (b) EW-control sample. 
The distribution for signal (VH),  shown for $m_{H}=125$~GeV, is multiplied 
by a factor of 100 and includes $ZH$ and $WH$ production.
The data are shown as points and the background contributions as histograms: 
dibosons are labeled as ``VV,'' ``V+l.f.'' includes $(W/Z)$+$(u,d,s,g)$ jets, 
``V+h.f.'' includes $(W/Z)$+$(b,c)$ jets and ``Top'' includes pair and single 
top quark production.}
\end{figure*}

\begin{figure*}[htp]
\centering
\subfigure[]{\includegraphics[width=8.5cm]{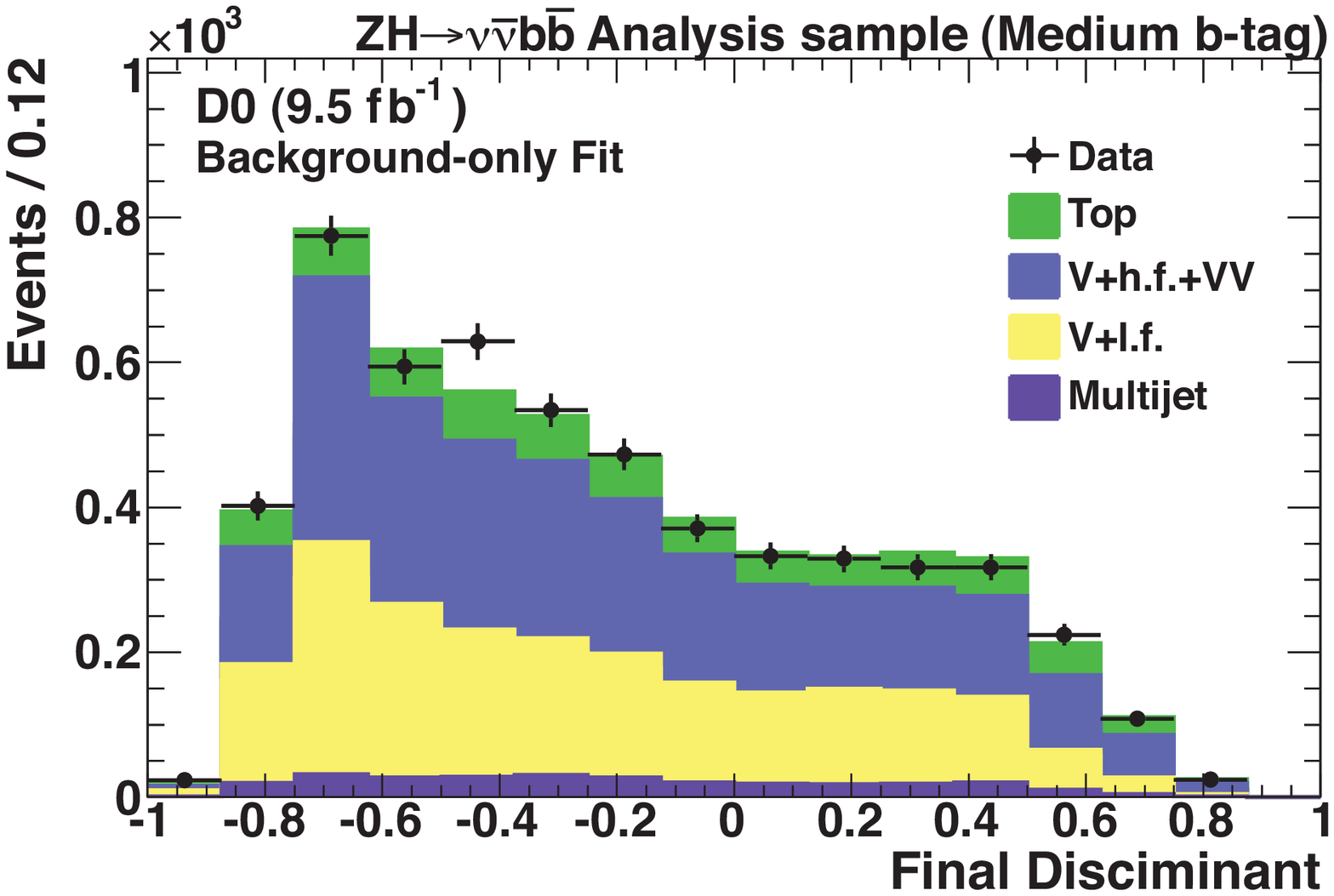}}
\subfigure[]{\includegraphics[width=8.5cm]{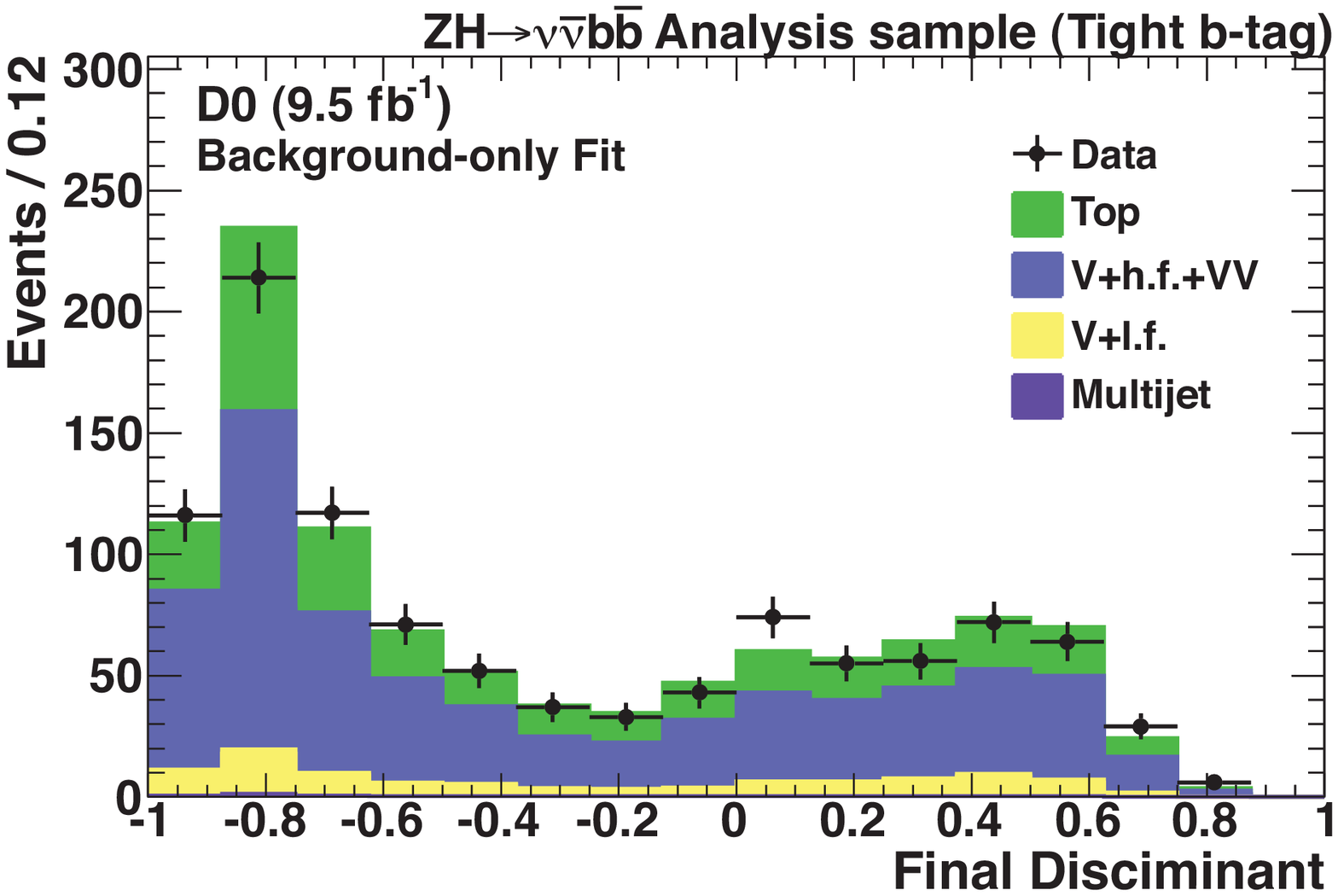}}
\caption{\label{bkgdsubtract}
The SM DT output, for $m_H=125$~GeV, following the multijet veto and
after the fit to the data under the background-only hypothesis in the
(a) medium and (b) tight $b$-tag channels. The data are shown as
points and the background contributions as histograms: dibosons are
labeled as ``VV'', ``V+l.f.'' includes $(W/Z)$+$(u,d,s,g)$ jets,
``V+h.f.'' includes $(W/Z)$+$(b,c)$ jets and ``Top'' includes pair and
single top quark production.}

\end{figure*}

\begin{figure*}[htp]
\centering
\subfigure[]{\includegraphics[width=8.5cm]{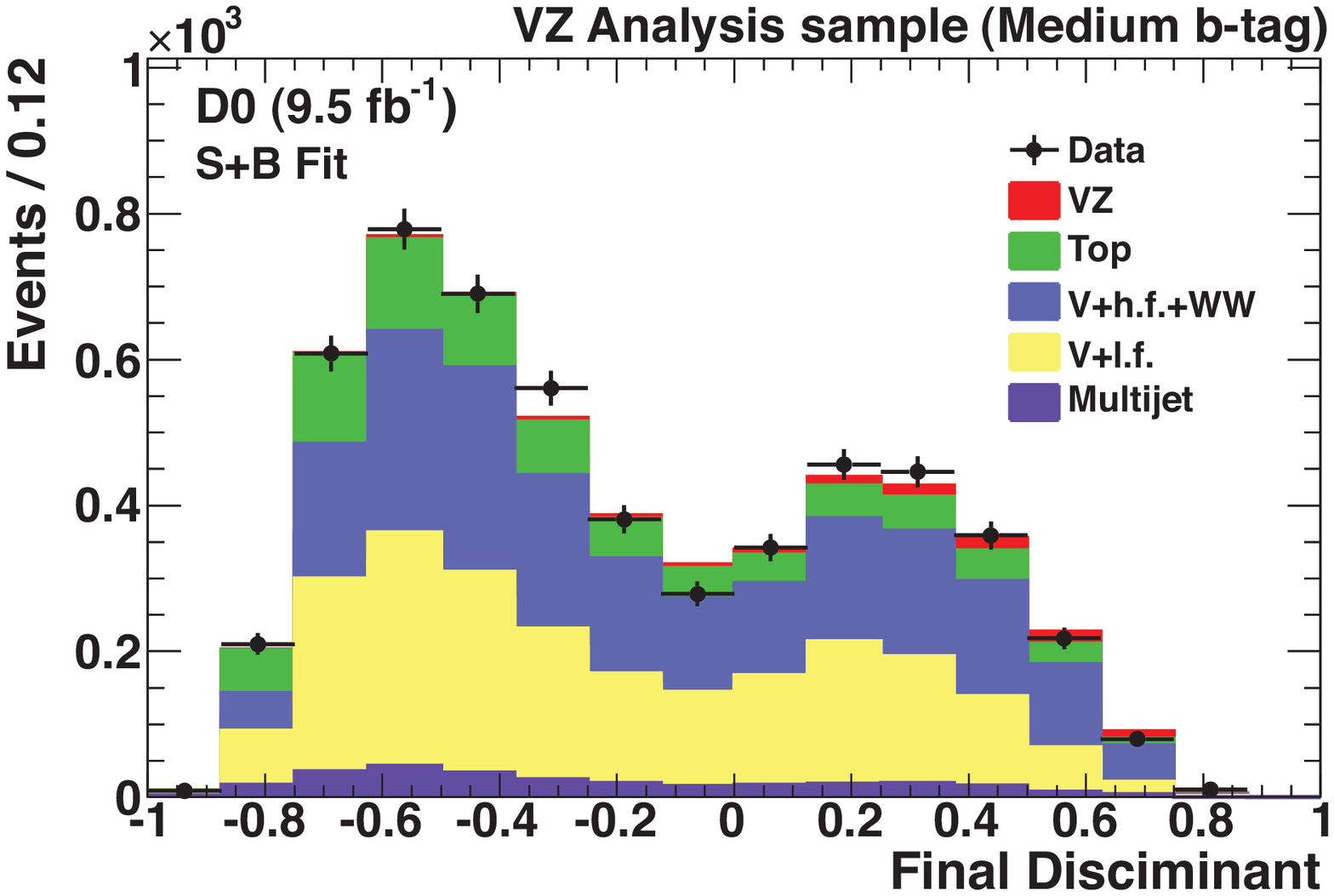}}
\subfigure[]{\includegraphics[width=8.5cm]{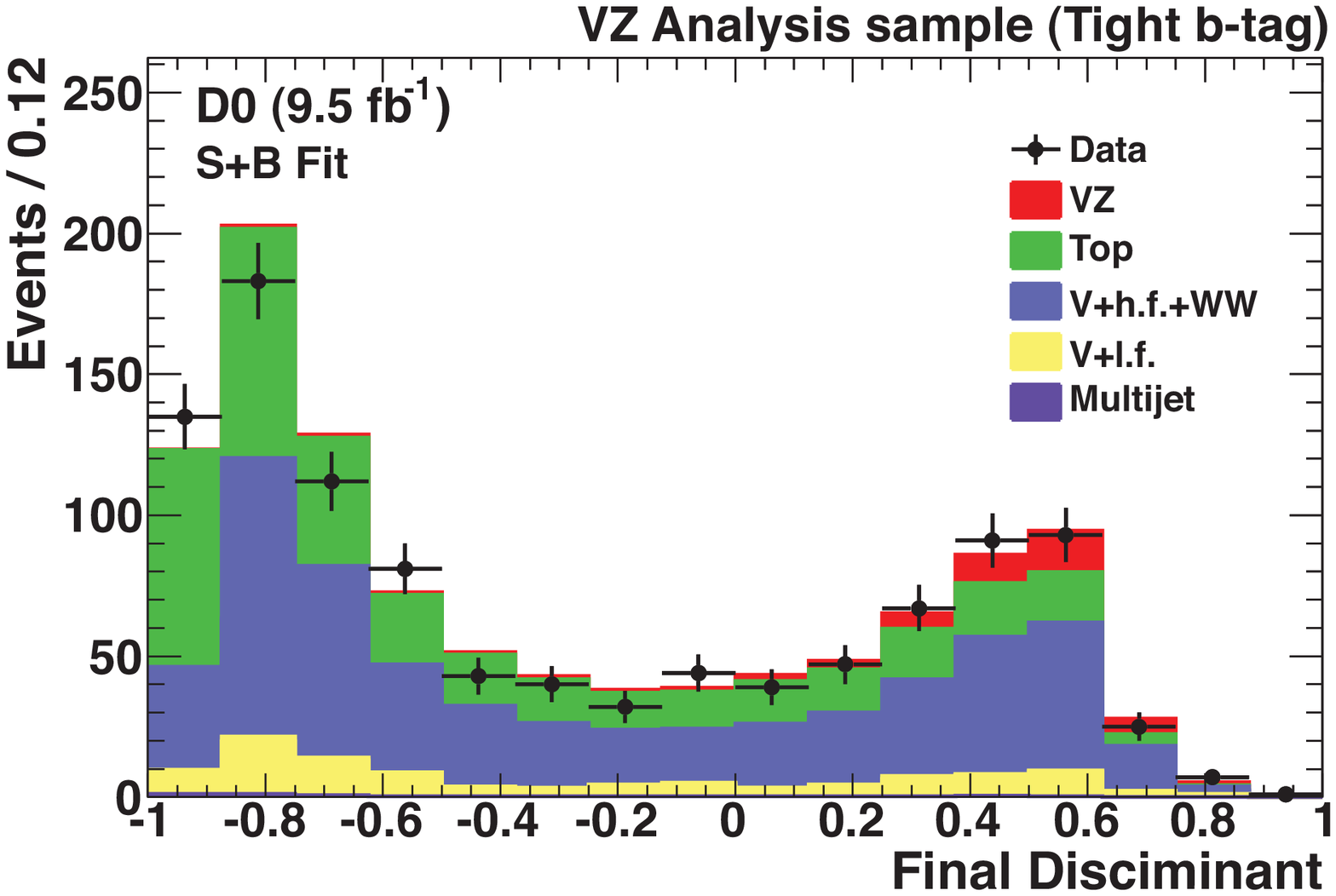}}
\caption{\label{bkgdsubtract_diboson_sbfit}
The SM DT output for the $WZ$ and $ZZ$ diboson search, 
following the multijet veto, 
and after the fit to the data under the signal+background hypothesis in the 
(a) medium and (b) tight tag channels. 
The data are shown as points and the background contributions as histograms; 
``V+l.f.'' includes $(W/Z)$+$(u,d,s,g)$ jets, ``V+h.f.'' 
includes $(W/Z)$+$(b,c)$ jets 
and ``Top'' includes pair and single top quark production.
The $WZ$ and $ZZ$ signal expectation (red histogram, and 
denoted VZ) is scaled to the SM cross section.
}
\end{figure*}

\end{document}